\documentclass[fleqn,usenatbib,useAMS]{mnras}
\pdfoutput=1

\usepackage[pdftex]{graphicx,color}
\usepackage[T1]{fontenc}
\usepackage[utf8]{inputenc}

\definecolor{urlblue}{rgb}{0,0,0.9}
\definecolor{linkblue}{rgb}{0,0,.8}
\definecolor{linkgreen}{rgb}{0,0.45,0}
\definecolor{linkpurple}{rgb}{0.7,0.0,0.4}
\definecolor{linkorange}{rgb}{0.7,0.1,0.0}

\usepackage{amsmath}	
\usepackage{amssymb}
\usepackage{amsbsy}
\usepackage{multicol}
\usepackage{multirow}
\usepackage{booktabs}
\usepackage{lmodern}

\usepackage{array}
\usepackage{graphicx} 
\usepackage{hhline}
\usepackage{makecell}

\usepackage{ae,aecompl}

\usepackage{newtxtext}
\usepackage{mathptmx}

\usepackage{listings}

\AtBeginDocument{\hypersetup{
linkcolor=linkblue,
citecolor=linkorange,
urlcolor=urlblue}}

\providecommand{\eprint}[1]{\href{http://arxiv.org/abs/#1}{#1}}
\providecommand{\adsurl}[1]{\href{#1}{ADS}}

\usepackage{ifthen}\def\eprinttmp@#1arXiv:#2 [#3]#4@{\ifthenelse{\equal{#3}{x}}{\href{http://arxiv.org/abs/#1}{#1}}{\href{http://arxiv.org/abs/#2}{arXiv:#2} [#3]}}
\renewcommand{\eprint}[1]{\eprinttmp@#1arXiv: [x]@}
\definecolor{DarkGreen}{rgb}{.1, .5, .1}
\newcommand{\tpot}{\textsc{TPOT}}
\newcommand{\deap}{\textsc{DEAP}}

\newcommand{\Var}{\mathrm{Var}}

\newcommand{\Loc}{\textbf{L16}}

\newcommand{\bm}[1]{\boldsymbol{#1}}

\usepackage{etoolbox}
\title{On the cosmological performance of photometrically classified supernovae with machine learning}

\author[MV dos Santos, M Quartin \& RRR Reis]{
Marcelo Vargas dos Santos,$^{1,2}$
Miguel Quartin,$^{1,3}$
Ribamar R. R. Reis$^{1,3}$
\\
$^{1}$Instituto de F\'isica, Universidade Federal do Rio de Janeiro, 21941-972, Rio de Janeiro, RJ, Brazil\\
$^{2}$Unidade Acadêmica de Física, Universidade Federal de Campina Grande, 58429-900 Campina Grande, PB, Brazil\\
$^{3}$Observat\'orio do Valongo, Universidade Federal do Rio de Janeiro, 20080-090, Rio de Janeiro, RJ, Brazil
}

\date{\today}

\pubyear{2019}

\begin{document}
\label{firstpage}
\pagerange{\pageref{firstpage}--\pageref{lastpage}}
\maketitle

\begin{abstract}
The efficient classification of different types of supernova is one of the most important problems for observational cosmology. However, spectroscopic confirmation of most objects in upcoming photometric surveys, such as the The Rubin Observatory Legacy Survey of Space and Time (LSST), will be unfeasible.  The development of automated classification processes based on photometry has thus become crucial. In this paper we investigate the performance of machine learning (ML) classification on the final cosmological constraints using simulated lightcurves from The Supernova Photometric Classification Challenge, released in 2010. We study the use of different feature sets for the lightcurves and many different ML pipelines based on either  decision tree ensembles or automated search processes.
To construct the final catalogs we propose a threshold selection method, by employing a \emph{Bias-Variance tradeoff}. This is  a very robust and efficient way to minimize the Mean Squared Error. With this method we were able to get very strong cosmological constraints, which allowed us to keep $\sim 75\%$ of the total information in the type Ia SNe when using the SALT2 feature set and $\sim 33\%$ for the other cases (based on either the Newling model or on standard wavelet decomposition).
\end{abstract}

\begin{keywords}
methods: data analysis; techniques: photometric; supernovae: general; cosmology: observations; astronomicaldatabases: miscellaneous;
surveys
\end{keywords}




\section{Introduction}

Current and next generation of optical surveys which cover large areas with good cadence, like the Dark Energy Survey~\citep[DES,][]{Abbott:2016ktf}, The Rubin Observatory Legacy Survey of Space and Time~\cite[LSST,][]{Abell:2009aa}, and the Zwicky Transient Facility \citep[ZTF,][]{Bellm:2014} will not provide spectra of the majority of observed objects. Alternatively, the Javalambre Physics of the Accelerating Universe (J-PAS) survey will provide effectively very coarse spectra with the use of 56 narrow band filters~\citep{Benitez:2014}, and other smaller surveys follow similar strategies, albeit with less filters~\citep{Cenarro:2019,Oliveira:2019}. This has pushed the community to develop tools which optimize the amount of science that can be gained from the purely photometric data. Most effort has been focused on how to determine the redshift (see, for instance, \cite{Abdalla:2008ze,Dahlen:2013fea,Benitez:2014,Sadeh:2015lsa,Gomes:2017plb} and references therein). For the cosmological use of supernovae (SNe), determining the redshift is not enough, as it is crucial to correctly classify them in order to isolate the type Ia SNe  (henceforth SNeIa, which are standardizable candles) from all other (core-collapse) types of SNe, for which distance estimations are not yet competitive when compared to the ones with SNeIa.

For the flat $w$CDM cosmological model, in which dark energy has a constant equation of state, taking into account all systematic uncertainties, SNeIa alone are able to constrain $\Omega_{m0}$ and $w$ to $23\%$ and $20\%$, respectively~\citep{Scolnic:2017caz}. And since SNe data imply different parameter degeneracies than CMB (Cosmic Microwave Background) and BAO (Baryon Acoustic Oscillations), the final uncertainties get much smaller when combining these datasets. A proper classification and redshift determination of future SNeIa would allow for unprecedent precision tests on the $\Lambda$CDM model as the number of events would grow from the current number of $\sim 10^3$~\citep{Betoule:2014frx,Scolnic:2017caz} to over $\sim 10^6$~\citep{Abell:2009aa}.

Such an increase in statistics opens up many new avenues for SN data exploration beyond the conventional Hubble-Lemaître diagram. For instance one can directly measure: anisotropies in the Hubble-Lemaître diagram~\citep{Koivisto:2010dr,Colin:2010ds,Soltis:2019ryf}; the Hubble constant for SNeIa in nearby galaxies~\citep{Riess:2016jrr,Burns:2018ggj}; non-Gaussianities in the scatter of the Hubble-Lemaître diagram due to lensing~\citep{Quartin:2013moa,Castro:2014oja,Macaulay:2016uwy}; correlations in the same scatter induced by SNeIa peculiar velocities~\citep{gordon2007,Castro:2015rrx,Howlett:2017asw,Garcia:2019ita}; the Hubble function combining SNeIa velocities and number densities~\citep{Amendola:2019lvy};
and time-delays between multiple images of strong lensed SNeIa~\citep{Oguri:2010ns,Bonvin:2016crt,Birrer:2018vtm,Huber:2019ljb}.

The SN classification scheme does not need to be boolean. Techniques have been proposed in~\citet{Kunz:2006ik,Newling:2011cp} to make use of SN catalogs for which each event has a set of probabilities of belonging to different types. Nevertheless, the higher the classification probability of each object, the better these kind of techniques can work. Current SN catalogs have employed a SN classification scheme based on an empirical template fitting~\citep{Sako:2014qmj,Jones:2017udy}. But in recent years alternative classifications scheme have been proposed~\citep{Ishida2012} and a couple of large classification challenges have been held~\citep{Kessler:2010qj,Malz:2018zlf}. 
In particular, the use of machine learning (ML) techniques in SN classification is starting to be studied in more detail~\citep{Lochner:2016hbn, Charnock:2016ifh, Moss:2018tug}. \cite{Ishida2019} indeed claims that the use of ML methods will be inevitable for the future of supernovae cosmology and automated transient classification.

The historical standard technique to perform photometric classification is  template matching. It consists in constructing the photometric light curves as a combination of spectra from a template set, and comparing it with the real one. The coefficients are then determined performing a numerical optimization (say, by a $\chi^2$ minimization); see for instance~\citet{Sako:2014qmj}. This is still the most common technique in Supernovae Surveys, and form the basis of SNe compilations such as the  SDSS-II~\citep{Kessler:2009ys}, JLA~\citep{Betoule:2014frx} and Pantheon~\citep{Scolnic:2017caz} catalogs. In this decade some groups started to use Machine Learning techniques to improve this classification. In SDSS-II the authors applied the Nearest Neighbors (NN) algorithm to see if it could improve their Photometric SN IDentification (PSNID) template fitting software~\citep{Sako:2014qmj}. The Pan-STARRS team \citep{Jones:2017udy} compared the NN with 3 different template fitting codes: PSNID, GalSNID~\citep{Foley2013} and Fitprob (the fit probability from the SALT2, Spectral Adaptive Lightcurve Template 2, light curve fit multiplied by a redshift-dependent SN type prior). Recently \citet{Villar:2019iuq} employed machine learning methods to photometrically classify Pan-STARRS SNe. The first public release of the \textsc{scikit-learn}\footnote{\url{scikit-learn.org}}~\citep{scikit-learn} library in 2010 increased the popularity of the ML methods.\footnote{\textsc{scikit-learn} provides sophisticated machine learning models in a  simple implementation, making it easy to write efficient codes with few lines.} 
In~\citet{Lochner:2016hbn} (henceforth \Loc) the authors performed the first thorough comparison between many \textsc{scikit-learn} models (including the NN) and found that an ensemble of decision-trees , the AdaBoost classifier, reached the best AUC score, which is the area under the \textit{Receiver Operating Characteristic} (ROC) curve (see below).

In this work we use the simulations performed as part of the Supernova Photometric Classification Challenge (SNPCC), discussed below in Section 3, in order to perform our ML training and tests. We start by performing a comparison similar to \Loc\ but expanding to other ensemble Decision Trees provided by \textsc{scikit-learn}: Adaptive Boosting (\textbf{ADA}), Random Forest (\textbf{RF}), Extra-Trees (\textbf{EXT}), Gradient Boosting (\textbf{GB}), and the Extreme Gradient Boosting (\textbf{XGB}) provided by the \textsc{XGBoost} library\footnote{\url{xgboost.ai}} and testing also an alternative metric to characterize the performance of the classifier: the \textit{Average Purity} (AP), while \Loc\ used only the AUC. We also performed an automated machine learning training by using \tpot\footnote{\url{epistasislab.github.io/tpot}} (Tree-Based Pipeline Optimization Tool), which allowed us to find the best pipelines for each feature set, while \Loc~used a simple pipeline for all cases, a \texttt{StandardScalar} transform followed by the machine learning model. We then explore how a ML-classified SN catalog performs in cosmological parameter estimation, quantifying the degradation due to the imperfect completeness and purity of the resulting catalogs. These results should be directly applicable to the SNe in DES due to the similar numbers of both SNe with only photometry and of SNe with spectra that can be used to train the algorithm.

This paper is structured as follows: in section \ref{sec:spcc} we summarize the SNPCC and its specifications; in section \ref{sec:ml} we revise the supervised learning approach paying special attention to the \emph{Bias-Variance tradeoff} method \ref{sec:bias_var}, with more details on decision trees \ref{sec:tree} and automated machine learning \ref{sec:automl}; in section \ref{sec:feat} discuss the different features sets. Finally we present our results and discuss them in section \ref{sec:results} and the conclusions in \ref{sec:conclusion}.

\section{Large photometric surveys and the Supernova Photometric Classification Challenge}\label{sec:spcc}

One of the greatest challenges in photometry is to infer values of observable quantities which are traditionally computed from spectra. The most recurrent example is the photometric determination of galaxy redshifts. In this work we focus solely on a separate photometric problem: supernova classification. While the different SN types are defined by the presence of specific spectral features, photometric classification must rely on their different lightcurve shapes and colors.

The Supernova Photometric Classification Challenge (SNPCC) was proposed by~\citet{Kessler:2010wk} in order to stimulate the development of tools to address this problem. They have publicly released a blinded mix of simulated SNe, with the different types selected in proportion to their expected rate and as if they had been measured with the \textit{griz} filters of the DES with realistic observing conditions (sky noise, point spread function and atmospheric transparency) based on years of recorded conditions at the DES site. Simulations of non-Ia SNe are based on spectroscopically confirmed light curves that include non-Ia samples donated by the Carnegie Supernova Project (CSP), the Supernova Legacy Survey (SNLS), and the Sloan Digital Sky Survey-II (SDSS–II).

The SNPCC catalog contains 21319 supernova light curves corresponding to five seasons of observations, simulated by the \textsc{SNANA} software. This set is composed by 5086 SNeIa and 16231 core collapse SNe, distributed among 7 different types: II, IIn, IIp, IIL, Ib, Ib/c and Ic. The type Ia supernovae were simulated using an equal mix of the models MLCS2k2~\citep{Jha:2006fm} and SALT2~\citep{Guy:2007dv} with an additional random color variation. An extinction correction, MLCS-U2 was used in order to make the models agree in the ultraviolet~\citep{Kessler:2009ys}. The non-Ia supernovae light curves were based on spectroscopically confirmed observed ones and constructed by warping a standard spectral template to match the photometry. The SN rates were based on the results from \citet{Dilday:2008wp} and \citet{Bazin:2009}. Ten groups addressed the challenge using different algorithms including template matching and ML techniques. The results of the challenge were published in \citet{Kessler:2010qj}: the template matching code PSNID from~\citet{Sako:2007ms} was found to be the best one based on the figure-of-merit employed (which was a function of purity and completeness).

After the end of the SNPCC the catalog was unblinded, and it now serves as a very useful tool to train and test ML algorithms relying on supervised training, in which a subset of the data has known classification \emph{a priori}. In our case in particular this corresponds to assuming we will have a subset of the SNe observed spectroscopically.

We discuss some technicalities involving the SALT2 lightcurve fitting analysis pertaining the simulated SNPCC supernovae in Appendix~\ref{app:salt2}.

\begin{figure*}
	\includegraphics[width=.47\textwidth]{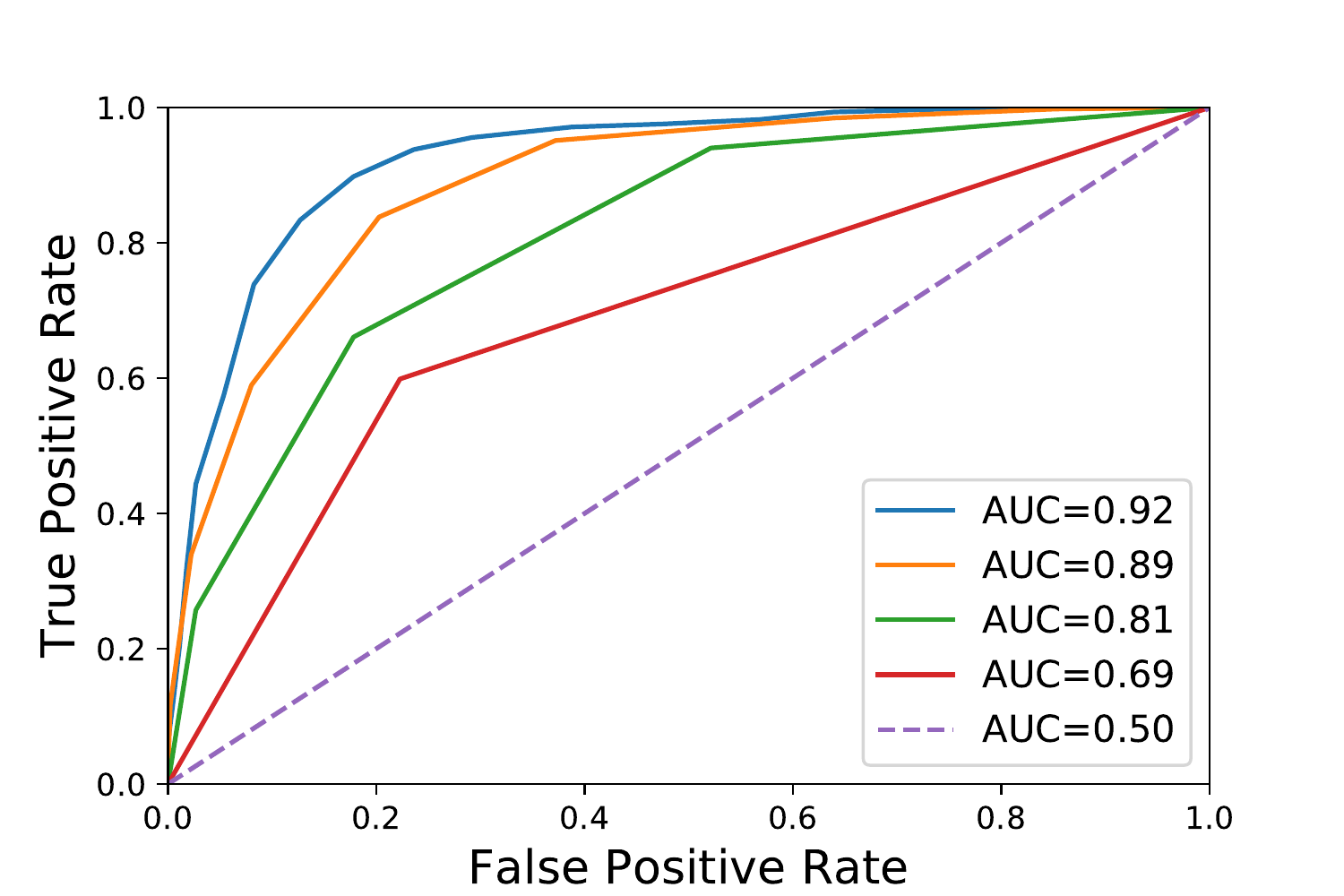}
	\includegraphics[width=.47\textwidth]{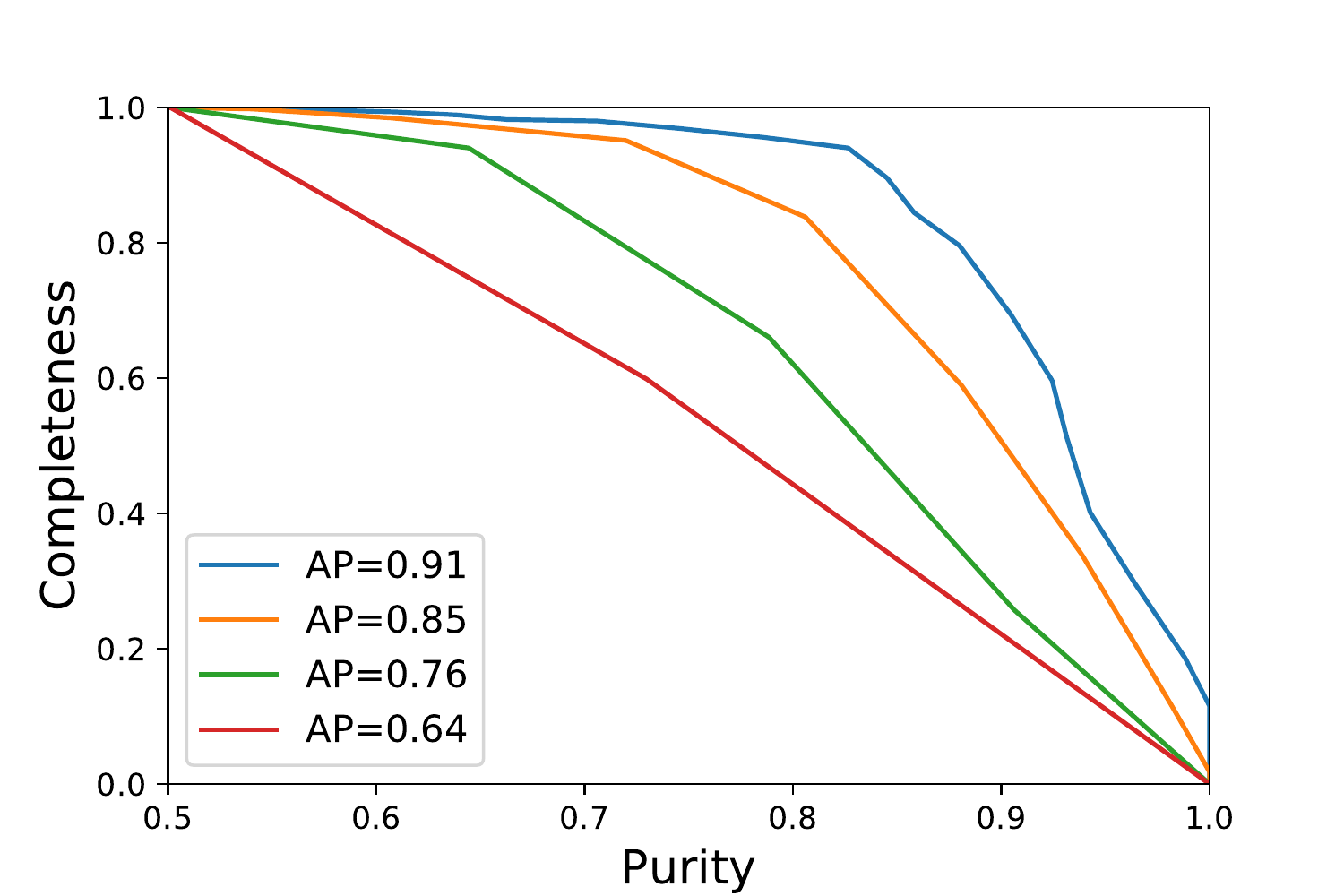}
\caption{Examples of ROC (left) and \textit{Completeness-Purity} (right) curves. In this example, the completeness reaches unity at a purity of 0.5, so we do not show the region below 0.5. Generally, the completeness can be below one for a minimum purity that can be less than 0.5.}
\label{fig:curves}
\end{figure*}

\section{The supervised machine learning approach}\label{sec:ml}

The objective of a supervised learning algorithm is to construct rules, according to a method, which relate a set of feature (observable) variables to a target one. If the rules are too simple it leads to \textit{underfitting}, which returns poor predictions and leads to poor performance. If the rules are instead too complex they may result in \textit{overfitting}, which results in a model that reproduces peculiar and unimportant behavior present in the training set data but not in the test data or on the real data.  A balance must be found and the hyperparameters (parameters of the model) chosen in order to optimize a score function. For a linear model, for instance, the hyperparameters are simply the intercept and the slope, usually set to minimize the mean squared error.

In classification problems, like the Supernovae classification, the target variables are classes, represented by integer numbers. In this work we are interested in the cosmological use of type Ia supernovae so we consider only two classes: Type Ia (Ia), the positive class (1); and Core Collapse (CC) supernovae, the negative class (0). The output of a classification is the probability that an input corresponds to a positive class, Type Ia in our case. In order to perform an univocal classification we must choose a \textit{probability threshold}, i.e. a probability value above which we assume to correspond to the positive class. The methods used to determine this value are referred to as the \textit{selection criteria}.

The quality of a classifier is completely summarized by four quantities: (i) The number of true positives (TP), i.e. the positive values which were correctly classified; (ii) The number of true negatives (TN), i.e. the negative values which were correctly classified; (iii) The number of false positives (FP), i.e. the positive values which were not correctly classified; (iv) The number of false negatives (FN), i.e. the negative values which were not correctly classified. These are the components of the confusion matrix, shown in Table~\ref{tab:confusion}.

All these quantities vary with the probability threshold, but always respect the constraint
\begin{equation}
    \rm{TP + FP + TN + FN = N},
\end{equation}
where $N$ is the length of the data set. So we must choose a value that maximizes the true predictions (TP and TN) and minimizes the false predictions (FP and FN). However we can not be sure that the best threshold found in the training process is the best for predictions. To circumvent this problem during the optimization process we choose a score function which depends of all threshold values (between 0 and 1). The score functions are derived from the confusion matrix, and in this work we use two different functions: the area under the \textit{receiver operating characteristic} (ROC) curve score, which is the same used in \Loc, and the Average Precision. To define these scores we must first define three other quantities.

\textit{Completeness} ($\mathcal{C}$) or \textit{True Positive Rate} (TPR): the fraction of real positives, which are correctly classified, used in final set. Also referred sometimes in the ML literature just as ``recall'' or ``sensitivity'', it is given by
\begin{equation}
    \cal{C}=\rm{TPR}=\frac{TP}{TP+FN} \,.
\end{equation}

\textit{False Positive Rate} (FPR): the fraction of real negatives which are not correctly classified, which contaminate the final set. Also referred as false alarm rate, it is given by
\begin{equation}
    \rm{FPR}=\frac{FP}{FP+TN} \,.
\end{equation}

\textit{Purity} ($\mathcal{P}$): The fraction of real positives in the final set. Sometimes referred to in the ML literature as just ``precision'', it is given by
\begin{equation}
    \cal{P}=\rm{\frac{TP}{TP+FP}} \,.
\end{equation}

\begin{table}
    \centering
    \renewcommand{\arraystretch}{1.6}
    \begin{tabular}{c c| c c}
    \multicolumn{1}{c}{} & \multicolumn{1}{c}{} & \multicolumn{2}{c}{True Class}\\
    \multicolumn{1}{c}{} & \multicolumn{1}{c}{} & \multicolumn{1}{|c}{P} & 
    \multicolumn{1}{c}{N}\\
    \cline{2-4}
    \multirow{2}{*}{Predicted class \hspace{.05cm}} & P& True positive (TP) & False positive 
    (FP)\\
    & N& False negative (FN) & True negative (TN)\\
    \end{tabular}
    \caption{\label{tab:confusion} Confusion matrix for a boolean classification problem.}
\end{table}

With the above definitions we may define the aforementioned scores.

\textit{Area Under the ROC Curve} (AUC): both TPR and FPR are, by definition, quantities between 0 and 1. They are used to construct the ROC curve, which is just a plot of TPR versus FPR, and is used as a metric for evaluating the performance of classification algorithms. Imperfect classifiers typically have monotonically increasing ROC curves, which often tend to zero in the limit in which FPR goes to zero and the area under this curve is naturally less than one, as shown in Figure \ref{fig:curves}. A perfect classifier has FN$=$FP$=0$ and thus ${\cal C}={\cal P}=1$ and TPR $=1$. Its ROC is technically ill-defined, but can be assumed to correspond to the limit in which the ROC tends to a step-function with TPR $=1$ and unity area.
The area under this curve is our first score function.

\textit{Average Purity} (AP): similarly we may define another curve as the plot of \textit{Purity} versus \textit{Completeness}. This curve is know as \textit{Completeness-Purity}~\citep{Markel:2019wll, Kgoadi2019, Villar:2019iuq}. In contrast with the ROC, this curve is decreasing. For the minimum threshold, where all objects are positively classified, the purity equates to the total fraction of the positives in the raw, unclassified data~\citep{Saito2015}. For higher thresholds the purity is expected to increase, so usually the minimum purity is simply
\begin{equation}
    \cal{P}_{\rm{min}} = \frac{\rm{TP+FN}}{\rm{N}}, \label{eq:Pmin}
\end{equation}
as shown in Figure \ref{fig:curves}. 
For a perfect classifier, the \textit{Purity} is equal to one and the AP curve corresponds to the unity square with ${\cal C}=1$, similar to the case of the ROC curve. However the area under this curve is not a reliable score thanks to the numerical noise. To avoid this problem we use the \textit{Average Purity} (dubbed Average Precision on \textsc{scikit-learn} and many ML references) as our second score function:
\begin{equation}
    \mathrm{AP}\equiv \sum_n (\mathcal{C}_{n} - \mathcal{C}_{n-1})\mathcal{P}_n,
\end{equation}
where the summation is over the thresholds.

Typically a ML problem consists in constructing an optimized pipeline of numerical preprocessing and machine learning methods that can give the best scores, where the score function is chosen according to the problem. As mentioned above, in this work we test the overall classification performance when optimizing one of two different score functions. The first, which follows the SNPCC approach, is the AUC -- see also~\cite{Swets-ROC-SciAm,fawcett2004roc} for accessible discussions on the subject. Since this metric is one of many which can be derived from the confusion matrix, we explore an alternative metric as well, which relies on the area under the curve of completeness versus purity.  This is motivated by the fact that in Astronomy one is usually more interested in high-purity catalogs. And since purity is not a linear combination of TPR and FPR, a code that maximizes this AP will not necessarily maximize the AUC. In addition, with the \textit{Completeness-Purity} curve we can easily infer the incompleteness of the final sample, which is important to manage possible biases in SN cosmology.

\begin{figure}
    \begin{center}
	    \includegraphics[width=.45\textwidth]{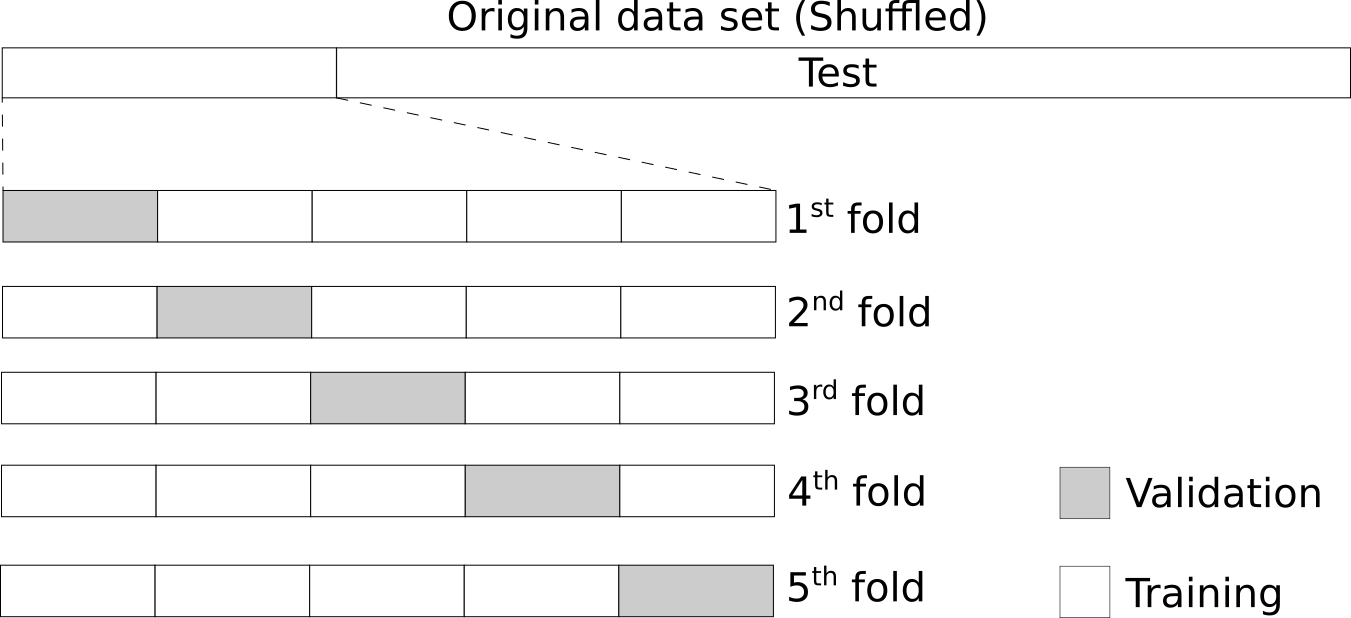}
	\end{center}
\caption{Flowchart depicting how the cross validation folding algorithm works.}
\label{fig:ML-flowchart}
\end{figure}

Our algorithm is depicted in Figure~\ref{fig:ML-flowchart}. It was constructed as follows: from the 21319 supernovae, we select randomly 1100 as our training set and leave the remaining 20219 as a testing set. We further subdivide the training set into 5 subsamples in order to perform a jack-knife approach: we use 4 out of 5 subsamples in the training process and cross-validate on the fifth subsample. This procedure is repeated for all 5 combinations internally by the \textsc{scikit-learn}  function \texttt{RandomizedSearchCV}, and the result with the best score in the validation set is kept.  This preliminary procedure is called 5-fold cross validation split in the ML literature, and is very important to avoid overfitting. It was also used by \Loc. Each training step consists in drawing a set of 30 points in the ML model hyperparameter space and finding which one maximizes the score in the testing set.

This whole procedure is repeated as many times as needed in order to get an accurate evaluation of the performance of the algorithm. In our case, we repeated this procedure at first 200 times for each ML method, each time using a different training set of 1100 supernovae, which is enough to get accurate results for the mean and variance of the AUC and AP. We later extended this number to 10000 repetitions for the best performing method in order to also get accurate results for the Bias-Variance tradeoff performance (discussed below).  

We also store a catalog of all SNeIa from each test set, which we identify as the result of a perfect classifier, i.e. the one which always returns $\mathcal{C}=1$ and $\mathcal{P}=1$. This corresponds to approximately 4800 SNeIa, for any probability threshold. This will be useful when we compare the cosmological performance of the photometric classified SNeIa with the ideal case in Section~\ref{sec:results_biasvar}.

\subsection{The \emph{Bias-Variance tradeoff}}
\label{sec:bias_var}

We define the bias as the average displacement of the estimated distance modulus from the fiducial value, $\Delta\mu=\bar{\mu} - \bar{\mu}_{\rm fid}(z_{\rm sim})$, and the Mean Squared Error (MSE) as the average squared displacement:
\begin{equation}
    \mathrm{b} = \langle \Delta\mu \rangle, ~~~ \mathrm{MSE} = \langle \Delta\mu^2 \rangle.
\end{equation}
We will also make use of the Root Mean Squared Error (RMSE) throughout the text. Note that there is a difference between two averages: the one denoted by a bar over the quantity (such as $\bar{\mu}$) represents the mean in a given bin of a given catalog, while the average denoted by $\langle ... \rangle$ represents the ensemble average over the 10,000 catalogs. Using the definition of variance
\begin{equation}
    \Var[\Delta\mu] = \langle \Delta\mu^2 \rangle - \langle \Delta\mu \rangle^2 = \mathrm{MSE} - \mathrm{b}^2,
\end{equation}
we find the relation between these quantities
\begin{equation}
    \mathrm{MSE} = \Var + \mathrm{b}^2.
\end{equation}

\begin{figure}
    \begin{center}
	    \includegraphics[width=.45\textwidth]{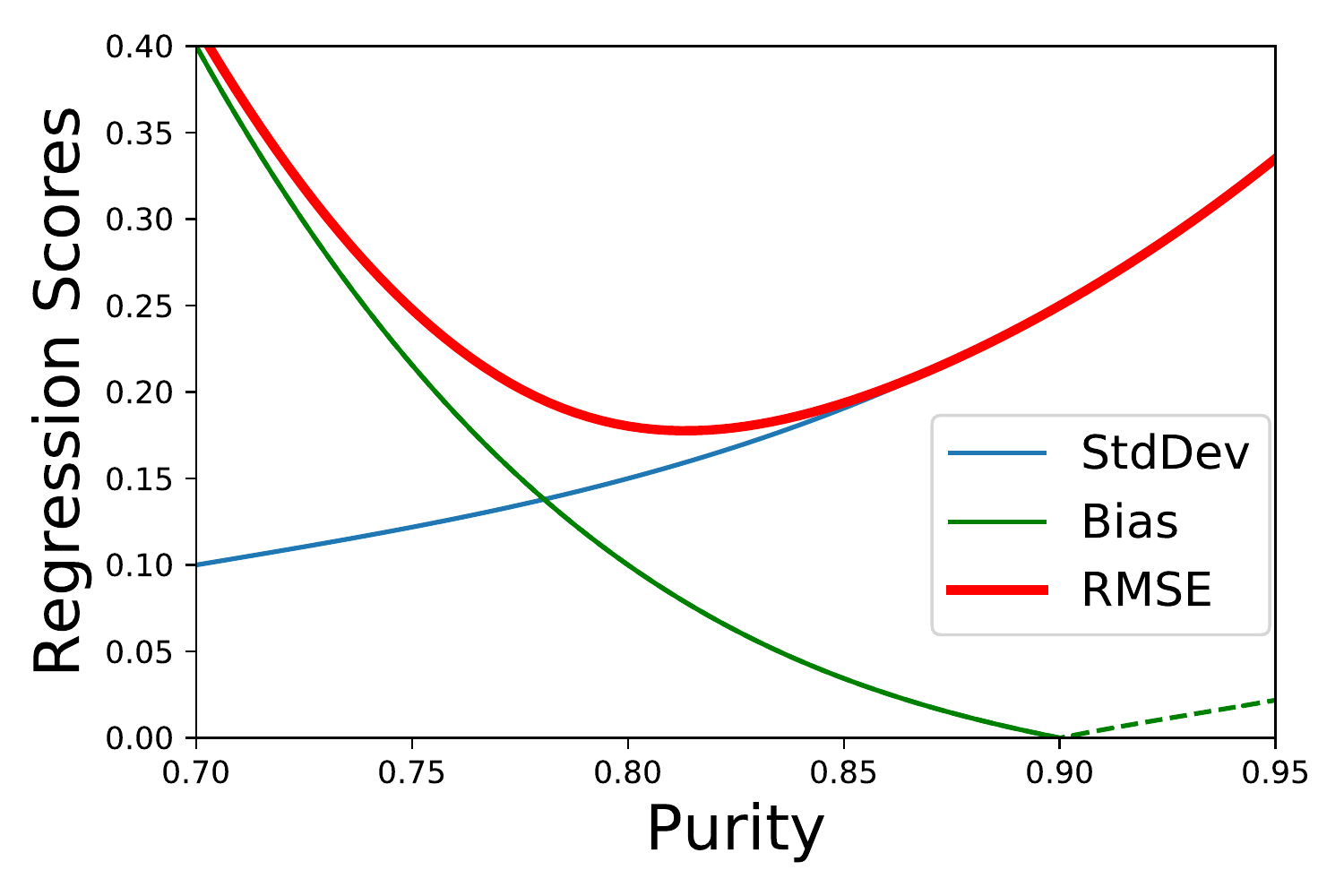}
	\end{center}
\caption{Illustration of a \emph{Bias-Variance tradeoff}. The minimum RMSE (red) depends on the balance between the bias $b$ (green) and standard deviation (blue). In the region where $b<0$ we show $|b|$ in dashed green. The fact that $b$ can be negative in our case substantially affects the optimization.}
\label{fig:bias_var}
\end{figure}

In general an increase in purity leads to a decrease in the bias at the cost of an increased variance. The Bias-Variance tradeoff consists in determining which hyperparameters values minimize the full MSE, instead of solely the variance. The typical \emph{Bias-Variance tradeoff}, as presented in most textbooks and tutorials, is enunciated for regression problems, where the balance between the variance and bias depends on the model complexity. However, since in this work we investigate how a classification problem affects the cosmological distance estimator, the tradeoff depends on the probability threshold, which can be mapped into the target purity of the catalog, as shown in Figure~\ref{fig:bias_var}. In this plot, the green line depicts the absolute value of the Bias and the dashed segment represents negative values. This sign flip can be very important for the optimization process, and in fact for our purposes it is, as we show in Section~\ref{sec:results_biasvar}.

\subsection{Decision trees and ensembles}\label{sec:tree}

\begin{figure*}
    \begin{minipage}{.67\textwidth}
    \begin{center}
	    \includegraphics[trim=1.5cm 1.4cm 1.5cm 1.5cm,clip=true,width=\columnwidth]{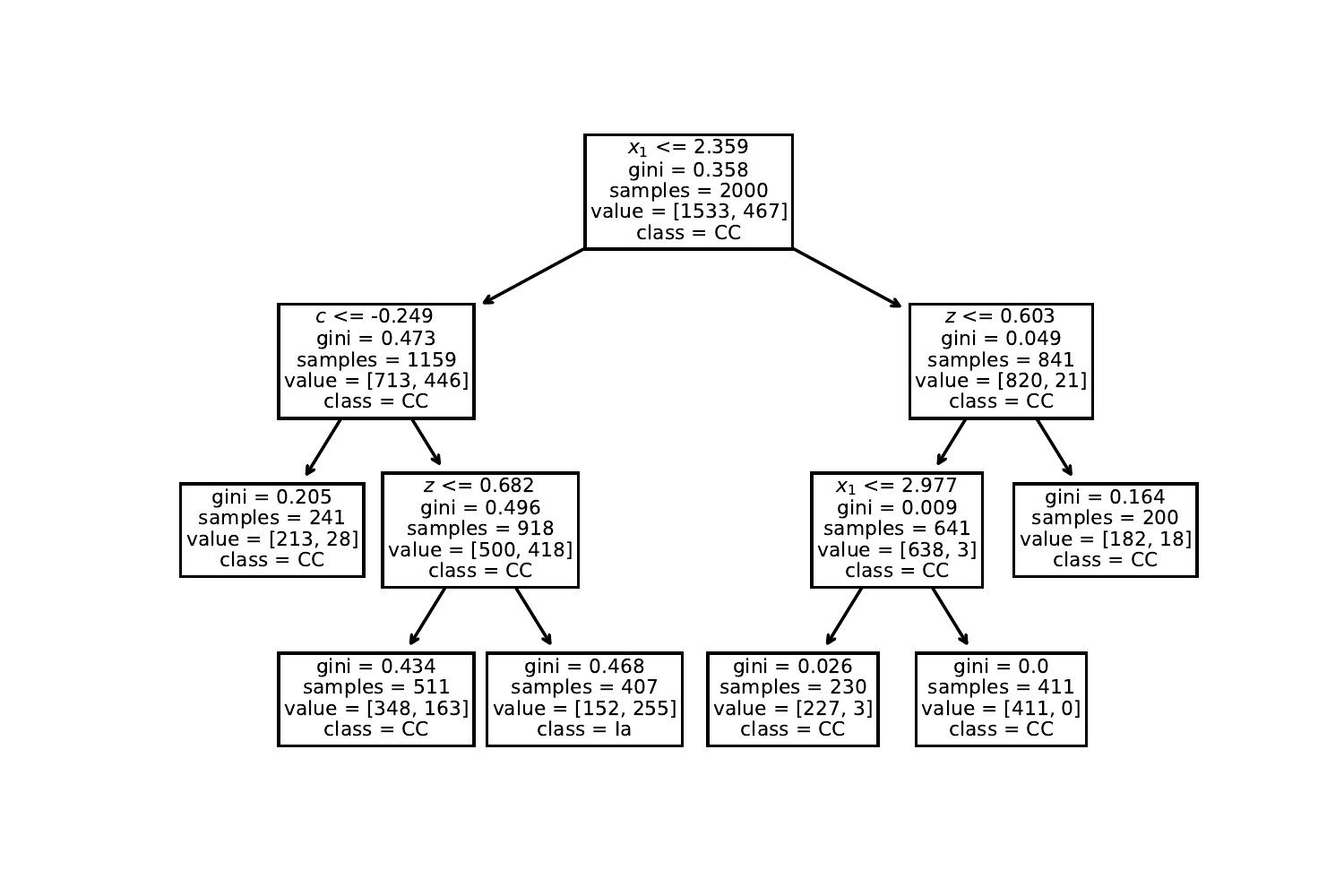}
	\end{center}
	\end{minipage}
\caption{Example of a decision tree generated by the \texttt{DecisionTreeClassifier}, function from \textsc{scikit-learn} library, trained on a 2000 supernovae sample using the SALT2 feature set. This graph is the output from the \texttt{export\_graphviz} function, from \texttt{sklearn.tree} module. The fist line (where present) is the condition for the feature; the second line is the quality criterion; 
the third is the number of samples from the training set used in the node; the fourth is a list of the number of samples for each class ([CC, Ia]); and the fifth is the class which corresponds to the test in the first line.}
\label{fig:tree}
\end{figure*}

The first approach that we have investigated is the ensemble of decision trees. The decision tree method consists in constructing a decision flow structure, known as tree, which splits the feature space in a rectangular grid. Each grid cell is then associated with a class (although a given class may end up mapped into various cells). Figure~\ref{fig:tree} shows an example of a small decision tree for the SALT2 feature set. The main parameters of a typical tree are the maximal depth and the minimum sample per leaf. The leaf is the final node, which returns a class.

An ensemble of trees is a combination of many different trees, randomly trained. These methods are very useful for analysis of light curves, and some of then were already investigated in recent papers~\citep{Lochner:2016hbn, Markel:2019wll, Kgoadi2019, Villar:2019iuq}. In this work we perform a broad comparison between five different ensemble methods. There are two different ways to build an ensemble of decision trees:
\begin{itemize}
\item[1.] \textit{Averaging Methods}: the driving principle is to build several independent weak estimators and then use the average of their predictions as the final prediction. On average, the ensemble estimator is usually better than any single estimator because its variance is reduced. In this work we use two such methods.
\begin{itemize}
    \item[i] Random Forest (\textbf{RF}), which constructs a distribution for the classes of the individual trees and take the mode of that as the final class.
    \item[ii] Extremely Randomized Trees or Extra-Trees (\textbf{EXT}), which includes a special kind of Decision Tree, known as Extra-Tree, that differs from classic decision trees in the way they are built. The thresholds are sorted randomly for each candidate feature and the ones with best results are used as the splitting rule.
\end{itemize}
\item[2.] \textit{Boosting Methods}: weak estimators, typically slightly better than a completely random choice, are built sequentially and one tries to reduce the bias of the combined estimator. In this work we used three different boost methods.
\begin{itemize}
    \item[i] Adaptive Boost, or AdaBoost (\textbf{ADA}), which is a weighted average of the individual predictions, from weak learners, where the weights are computed iteratively.
    \item[ii] Gradient Boost (\textbf{GB}), which uses the gradient descendent method to minimize the loss function.
    \item[iii] Extreme Gradient Boost(\textbf{XGB}), which is an optimized implementation of the \textbf{GB} available in \texttt{xgboost.ai}.
\end{itemize}
\end{itemize}

\subsection{Automated Machine Learning}\label{sec:automl}

The automated Machine Learning approach aims to determine what is the best pipeline to train the machine learning model by an automated process, i.e. a code that can test several pipelines and assess which one is the best for the problem. The \tpot (Tree-based Pipeline Optimization Tool)\ uses the evolutionary computation framework provided by the \deap\ (Distributed Evolutionary Algorithms in Python)\footnote{\url{deap.readthedocs.io}} library to choose the best machine learning pipeline.

A typical \tpot\ pipeline is composed by three main steps: (i)~a combination of one or more transformed copies of the features; (ii) a recursive feature elimination; and (iii) a classifier (or a regressor), which could be a single method or a stacking of many.

One of the most simple pipelines in our work was found when we optimized a classifier for Karpenka features, with AUC as metric:
\begin{lstlisting}
GradientBoostingClassifier(
    learning_rate=0.1, max_depth=8,
    max_features=0.85, min_samples_leaf=4,
    min_samples_split=11, n_estimators=100,
    subsample=0.9
)
\end{lstlisting}
This means that  steps (i) and (ii) are reduced to the identity, and (iii) becomes a single \texttt{GradientBoostingClassifier}. But when we include $z_{\rm host}$ as a new feature we find a completely different pipeline:
\begin{lstlisting}
make_union(
    FunctionTransformer(copy),
    make_pipeline(
        VarianceThreshold(threshold=0.2),
        FastICA(tol=0.0)
    )
),
XGBClassifier(
    learning_rate=0.1, max_depth=9,
    min_child_weight=2, n_estimators=100,
    nthread=1, subsample=0.8
)
\end{lstlisting}
Step (i) is now to take two copies of the input data and transform one applying a \texttt{VarianceThreshold} and \texttt{FastICA} in this order,  step (ii) is a union of these two copies and (iii) is a \texttt{XGBClassifier}. All the \tpot pipelines are shown in Appendix~\ref{app:tpot}.

\section{Features Extraction}\label{sec:feat}

\subsection{Template fitting}\label{sec:template}

The template fitting\footnote{Note that \emph{Template Fitting}  refers to methods which use templates of Type Ia light curves to fit other light curves, while \emph{Template Matching} refers to ones which use templates of different types to classify any light curve.} approach consists in reconstruct the light curve of a type Ia supernova, based on templates constructed from previous data sets of SNeIa confirmed by spectroscopic surveys. We fit the light curves computing the flux of the SALT2 model by using the implementation provided by the \textsc{SNCosmo}\footnote{\url{sncosmo.readthedocs.io}} library (\url{sncosmo.readthedocs.io}), and sample 1000 points using the multinest method provided by the \textsc{PyMultinest}\footnote{\url{github.com/JohannesBuchner/PyMultiNest}} python library.

In SALT2, the rest-frame specific flux at wavelength $\lambda$ and phase (time) $p$ is
modeled by ($p = 0$ at $B$-band maximum) 
\begin{equation}
\phi(p,\lambda; x_0,x_1,c)=x_0[M_0(p,\lambda)+x_1M_1(p,\lambda)]
\exp [c \, CL(\lambda)],
\label{eq:salt2restflux}
\end{equation}
where $M_0(t,\lambda)$ is the average spectral sequence (using past supernova data, \cite{Guy:2007dv}), and $M_k(t,\lambda)$, $k > 0$, are higher order components related to the supernova variability (terms with $k>1$ are negligible). Finally, $CL(\lambda)$ is the average color correction.

The
observer-frame
flux in passband $Y$ is calculated as
\begin{equation}
F^Y(p(1+z))=(1+z)\int \phi(p,\lambda')T^Y(\lambda' (1+z)) d\lambda',
\label{eq:salt2obsflux}
\end{equation}
where $T^Y(\lambda)$ is the transmission curve of the observer-frame
filter $Y$, and $z$ is the redshift. The free parameters to be fitted are $x_0$, $x_1$ and $c$ which are the SED sequence normalization, the stretch and the color parameters respectively, and also $t_0$ (the time of B band maximum), and the supernova redshift $z$. So in total we have 5 free parameters, for which we use the priors given in Table~\ref{tab:salt2}.
{
\renewcommand{\arraystretch}{1.5}
\begin{table}
\centering
\begin{tabular}{cc}
\hline
 Parameter & Range\\
\hline
  $z$ & $(0.01,1.5)$\\
  $t_0$ & $(-60,100)$\\
  $x_0$ & $(-10^{-3},10^{-3})$\\
  $x_1$ & $(-3,3)$\\
  $c$ & $(-0.5,0.5)$\\
\hline
 \end{tabular}
 \caption{Uniform prior ranges on the SALT2 model parameters. \label{tab:salt2}}
 \vspace{6pt}
\end{table}
}

\subsection{Parametric models}\label{sec:parametric}

This approach consists in fitting a functional shape for the light curves for each filter. The advantage over the previous method is that we make less prior assumptions about supernovae. In this work we use the fitted parameters as features for our machine learning algorithms. To fit the light curves we use two models proposed for the SNPCC~\citep{Kessler:2010qj, Kessler:2010wk}: (a) The Newling model~\citep{Newling2010}; (b) the Karpenka model~\citep{Karpenka2012}. The parameters for both parametric models were determined by sampling 1000 points with \textsc{PyMultiNest}.

\subsubsection{Newling model}\label{sec:newling}

The function form proposed by the Newling model is given by
\begin{equation}
    F(t) = A \left(\frac{t-\phi}{\sigma} \right)^k \exp \left(- \frac{t-\phi}{\sigma} \right) 
    k^{-k} e^k + \Psi(t),
\end{equation}
where 
\begin{equation}
    \Psi(t) = 
    \begin{cases}
    0 &-\infty < t < \phi\\
    \text{cubic spline} &\hphantom{\infty}\phi < t < \tau \\
    \psi &\hphantom{\infty}\tau < t < \infty \\
    \end{cases}. \label{eq:Psi}
\end{equation}
The peak flux is equal to $A+\psi$, $\phi$ is the explosion starting time, $k$ is related to the relative rise and decay times and $\sigma$ is the light curve width, which gives 5 free parameters. The $\Psi(t)$ is a tail function which ensures that the flux tends to a finite value for $t\gg\tau$, where $\tau = \phi+k\sigma$ is the peak time. The cubic spline in equation~\eqref{eq:Psi} is determined to null the derivative in $t=\phi$ and $t=\tau$. The model is fitted separately for each filter, giving a total of 20 features, for the SDSS $griz$ filter system used in the SNPCC. The parameters ranges (which are the same for all bands) are given in Table~\ref{tab:newling}.

{ 
\renewcommand{\arraystretch}{1.5}
\begin{table}
\centering
 \begin{tabular}{cc}
 \hline
  Parameter & Range\\
 \hline
  $\log(A)$ & $(0,10)$\\
  $\phi$ & $(-60,100)$\\
  $\log(\sigma)$ & $(-3,4)$\\
  $\log(k)$ & $(-4,4)$\\
  $\log(\psi)$ & $(-6,10)$\\
  \hline
 \end{tabular}
\caption{Uniform prior ranges on the Newling model parameters.}
\label{tab:newling}
\end{table}
}

{ 
\renewcommand{\arraystretch}{1.5}
\begin{table}
\centering
 \begin{tabular}{cc}
 \hline
  Parameter & Range\\
 \hline
  $\log(A)$ & $(-3, 3)$\\
  $\log(B)$ & $(-3, 2)$\\
  $t_0$ & $(0,100)$\\
  $t_1$ & $(0,100)$\\
  $T_{\text{rise}}$ & $(0,100)$\\
  $T_{\text{fall}}$ & $(0,100)$\\
  \hline
 \end{tabular}
\caption{Uniform prior ranges on the Karpenka model parameters. \label{tab:karpenka_priors}} 
\end{table}
}

\subsubsection{Karpenka model}\label{sec:karpenka}
The Karpenka model is similar to Newling's, generating a light curve with a peak and a tail which tends to a finite value, but can also describe a situation where a second flux peak can happen. The function is given by
\begin{equation}
 F(t) = A \big[1+B(t-t_1)^2 \big] \, 
\frac{\exp[-(t-t_0)/T_\text{fall}]}{1+\exp[-(t-t_0)/T_\text{rise}]},
\end{equation}
where $t=0$ is taken as the earliest measurement in the r-band light curve. There are 6 free parameters $\{A, B, t_0, t_1, T_\text{fall}, T_\text{rise}\}$ for each filter, totalizing 24 free parameters. The ranges of the priors are shown in Table~\ref{tab:karpenka_priors}, which are the same for all bands.

\subsection{Wavelet decomposition}\label{sec:wav}

The wavelet decomposition is a very useful tool for signal processing and time series analysis, which is the case of supernovae light curves. Following \Loc~we perform the \textit{\`a trous} wavelet transform, which achieves dyadic scale-invariance, and use the symlet family of wavelets, which are a more symmetric version of the widely used Daubechies family of wavelets, by the implementation \texttt{sym2} family provided by the \textsc{PyWavelets} library.\footnote{\url{pywavelets.readthedocs.io}} Before using the wavelet decomposition we construct an interpolated version of the light curve, for each band, using the Gaussian Random Process provided by the \textsc{George} library.\footnote{We choose the exponential radial kernel optimized following this procedure: \url{george.readthedocs.io/en/latest/tutorials/hyper}}

\begin{figure*}
	\includegraphics[width=.74\textwidth]{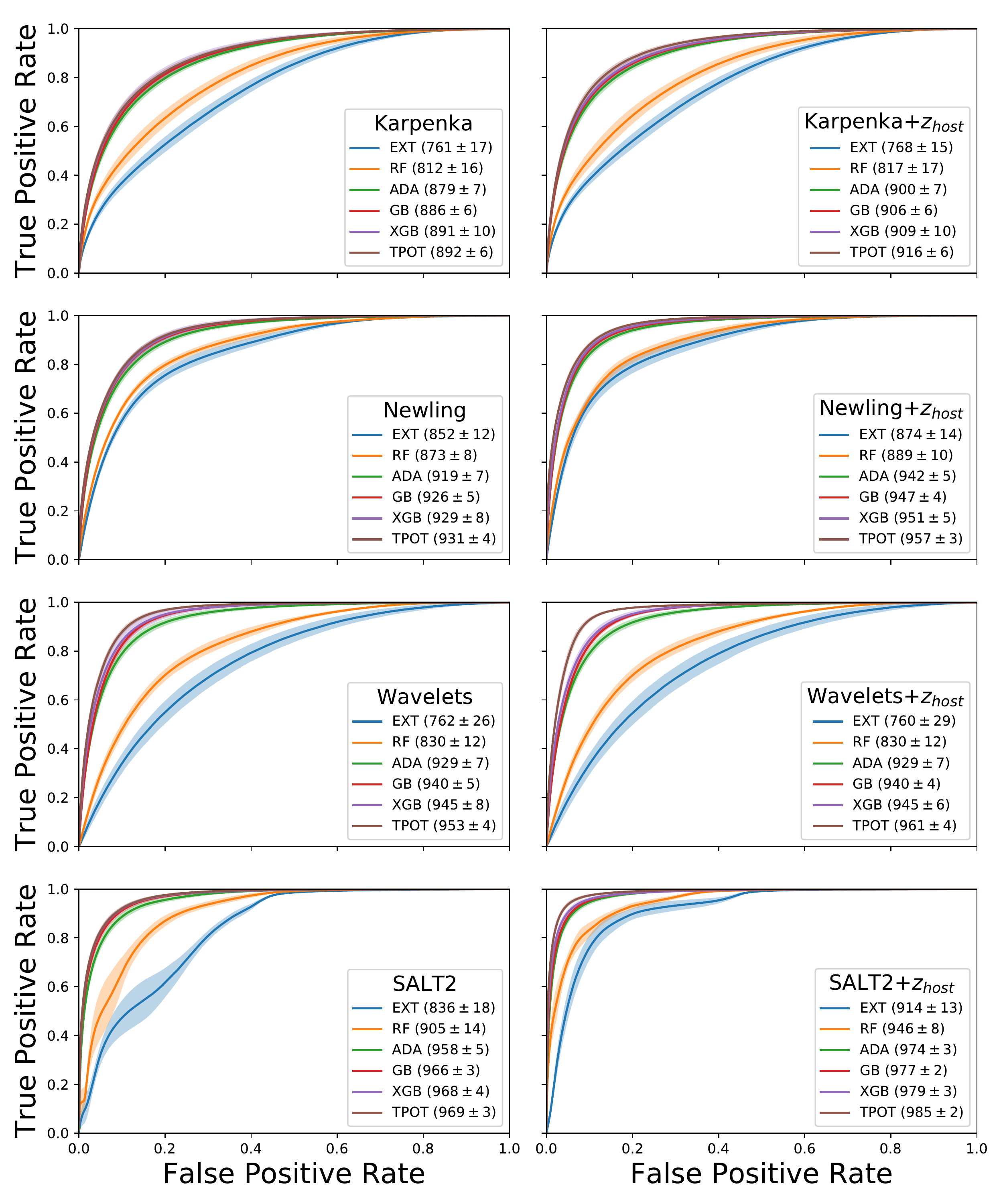}
\caption{ROC curves for each feature set, without (left) and with (right) the host galaxy redshift information, for all ML techniques. The numbers in the legends correspond to the value of $1000\times$AUC (the area under the curve) for each case.}
\label{fig:rocs}
\end{figure*}

We choose 100 points, evenly spaced in the time range of the light curve, for the Gaussian process curve and a two-level wavelet transform, returning 400 coefficients per filter, or 1600 coefficients per object. To avoid redundancy we perform a dimensionality reduction with Principal Components Analysis (PCA) restricting to only the first 20 components, which we use as our feature set and refer only by Wavelets instead Wavelet decomposition.

\subsection{Host galaxy photometric redshift ($z_{\rm host}$) as a new feature}

Following the SNPCC we also investigated the effects of introducing the host galaxy photometric redshift, $z_{\rm host}$, as a new feature. So for each feature set we performed the classification twice, both with and without $z_{\rm host}$ in the feature set. For Wavelets, for instance, we first train using the features $({\rm PCA}_1, {\rm PCA}_2, ..., {\rm PCA}_{20})$, and then a second time using  $({\rm PCA}_1, {\rm PCA}_2, ..., {\rm PCA}_{20}, z_{\rm host})$. The only exception is the SALT2 case, where the features are the best fit parameters of two different light curve fits and the redshift can simply be a free parameter in the fit. First, we let the redshift be free and determined only by the light curve, thus using the feature set $(z, t_0, x_0, x_1, c)$. To include $z_{\rm host}$ we refit the light curve fixing $z=z_{\rm host}$ and fitting $(t_0, x_0, x_1, c)$ again, in which case the feature set becomes $(z_{\rm host}, t_0, x_0, x_1, c)$.

The value of $z_{\rm host}$ is the photometric one, both for training and predicting. We have tested using the true spectroscopy redshift to train both the models and the photometric prediction. But this resulted in smaller scores than the ones shown below. Even though $z_{\rm host}$ is not exactly the spectroscopic one, the latter will not be measurable. This downgrade in performance when using the exact redshift is likely due to the fact than in ML methods, the closer your training set is to the real data, the better the performance.

\begin{figure*}
	\includegraphics[width=.74\textwidth]{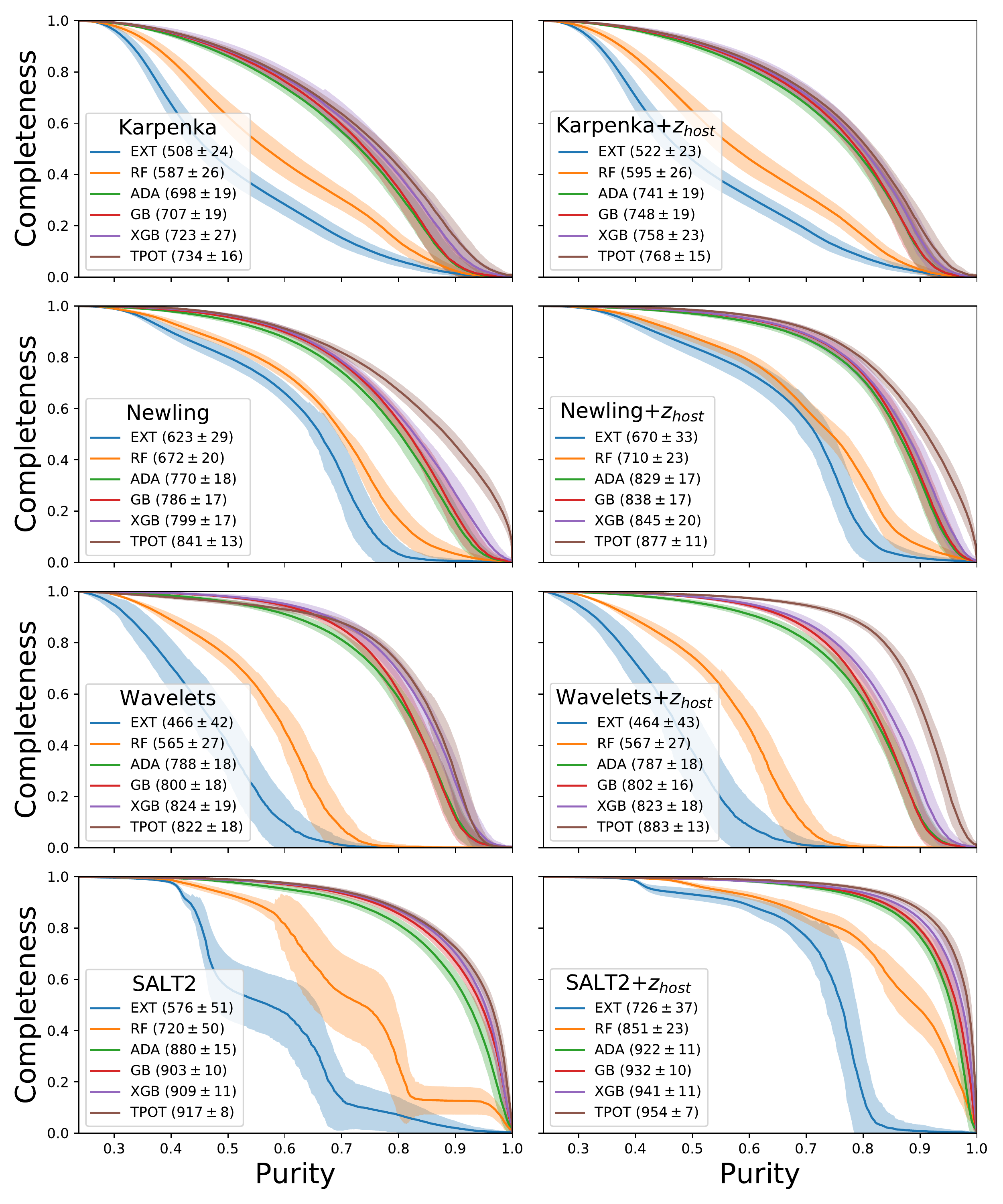}
\caption{Same as Figure~\ref{fig:rocs} for the Completeness-Purity curves. The numbers in the legends correspond here to the value of $1000\times$AP for each case.}
\label{fig:prcs}
\end{figure*}

\section{Results and discussion}\label{sec:results}

\subsection{Comparison of ML methods with AUC and AP}\label{sec:results_ml}

\begin{figure*}
	\includegraphics[width=0.77\textwidth]{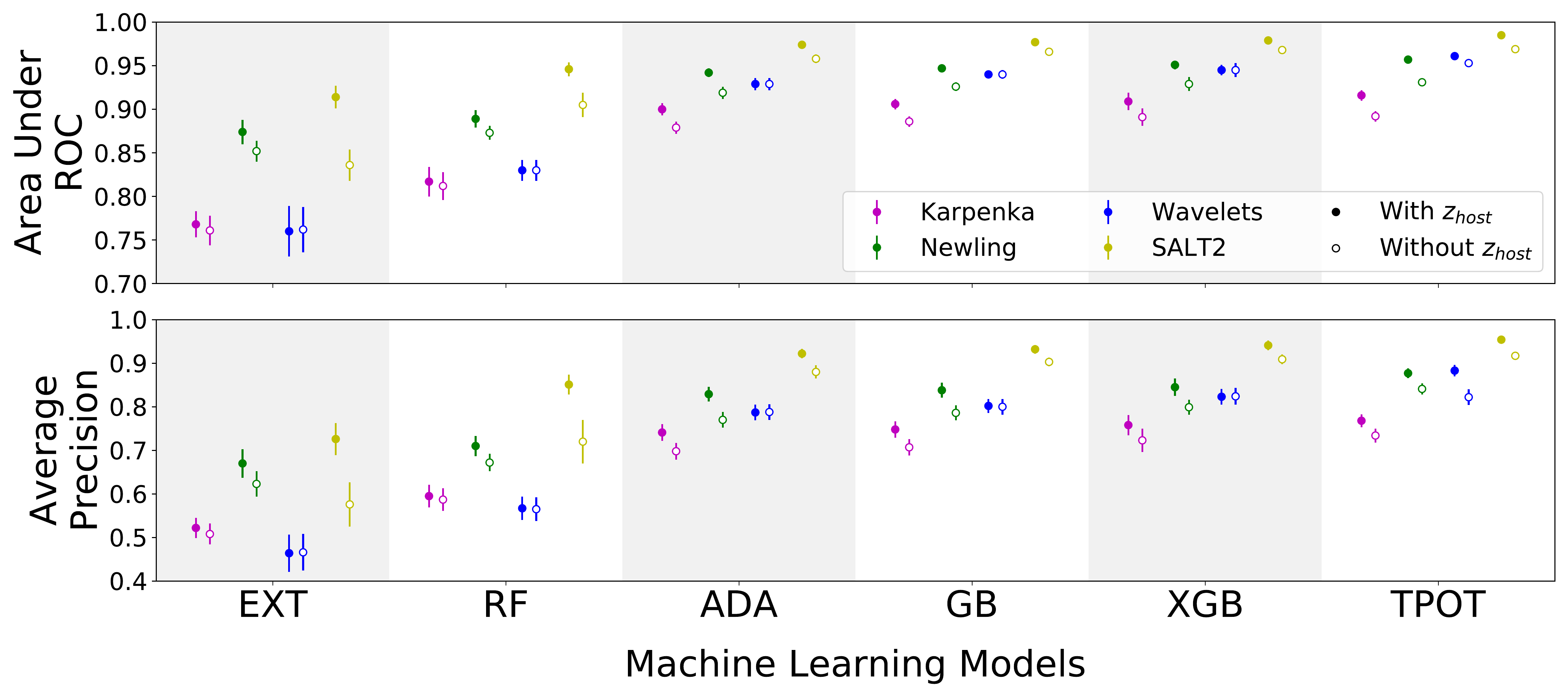}
\caption{Comparison between AUC and AP for each feature set and ML method.}
\label{fig:scores}
\end{figure*}

Let us first analyze the results of the decision trees ensembles. In Figure~\ref{fig:rocs} we compare the \textrm{ROC} curves and the AUC value (given in \emph{per mille} in the plots) for each ML model and feature set. The curves and shaded regions around them are the mean and the standard deviation in the catalog set. The AUC values and  uncertainties are likewise obtained from this set. 
We notice that the boosting methods (ADA, GB, and XGB) reach higher scores than the averaging methods (EXT and RF) for any set of features. It is also clear that the different boosting methods exhibit very similar performance for all features. \tpot~was not able to provide a significant improvement, except for two cases, wavelets and SALT2 in the case in which we include the $z_{\rm host}$. The \textit{Completeness-Purity} curves are in turn depicted in Figure~\ref{fig:prcs}. We note in general the same difference between boosting and averaging methods, although the latter exhibit broader distributions. An interesting feature which is made clear in this score is that even methods with high AP may result in considerably low completeness values for high purities. This happens in general for Wavelets, Karpenka and Newling, but in some cases using \tpot~this issue was alleviated. This means that we may lose a lot of candidates if we need to impose a restrictive purity threshold, which in turn increases the statistical error. For Wavelets (ADA, GB, XGB), for example, if we impose a 90\% threshold for the purity, we get a catalog with less than $10\%$ completeness. In contrast, for SALT2, we get around 70\% of the real positives even when we impose a 90\% purity.

Figure~\ref{fig:scores} compiles the scores for all cases, where we can confirm the ranking between the methods described above, and also compare the features more clearly. The ranking, from low to high score, is: Karpenka, Newling, Wavelets and SALT2. We can also confirm that the introduction of $z_{\rm host}$ as feature increases the score as expected and already reported in \Loc, except for the case of Wavelets in which it only made any difference when we used \tpot.

Using AP as score, we found that the \tpot\ method was able to increase significantly the scores, as can be seen clearly in Figure~\ref{fig:prcs}. It was particularly successful for the Newling model and,  when $z_{\rm host}$ is included, and also for the Wavelets method.  Summarizing the scores for all cases, we find the following.
\begin{itemize}
    \item Comparing the feature set for different methods we find from best to worst: 1. SALT2; 2. Newling and Wavelets (tied); 3. Karpenka (although when using the averaging ML methods Newling performed better). We stress that since the SALT2 model was used to create the lightcurves the results may be biased toward it. Nevertheless as we discuss below the performance of Newling and Wavelets are still very competitive. \\
    \item  Comparing the ML methods for given feature sets, from best to worst: 1. \tpot; 2. XGB; 3. GB; 4. ADA; 5. RF; 6. EXT. \\
    \item Including $z_{\rm host}$ as a new feature: the scores increased in general, as expected, except for Wavelets where it only increased when using \tpot.
    However, one should note that the increase in performance obtained by including the host redshift was comparable to the performance difference among ML methods and feature sets. Therefore using \tpot\ without redshift information still yielded very good results, which is a promising result given that future photometric surveys will only have limited spectroscopic follow-ups.
\end{itemize}

\begin{figure*}
    \includegraphics[width=0.75\textwidth]{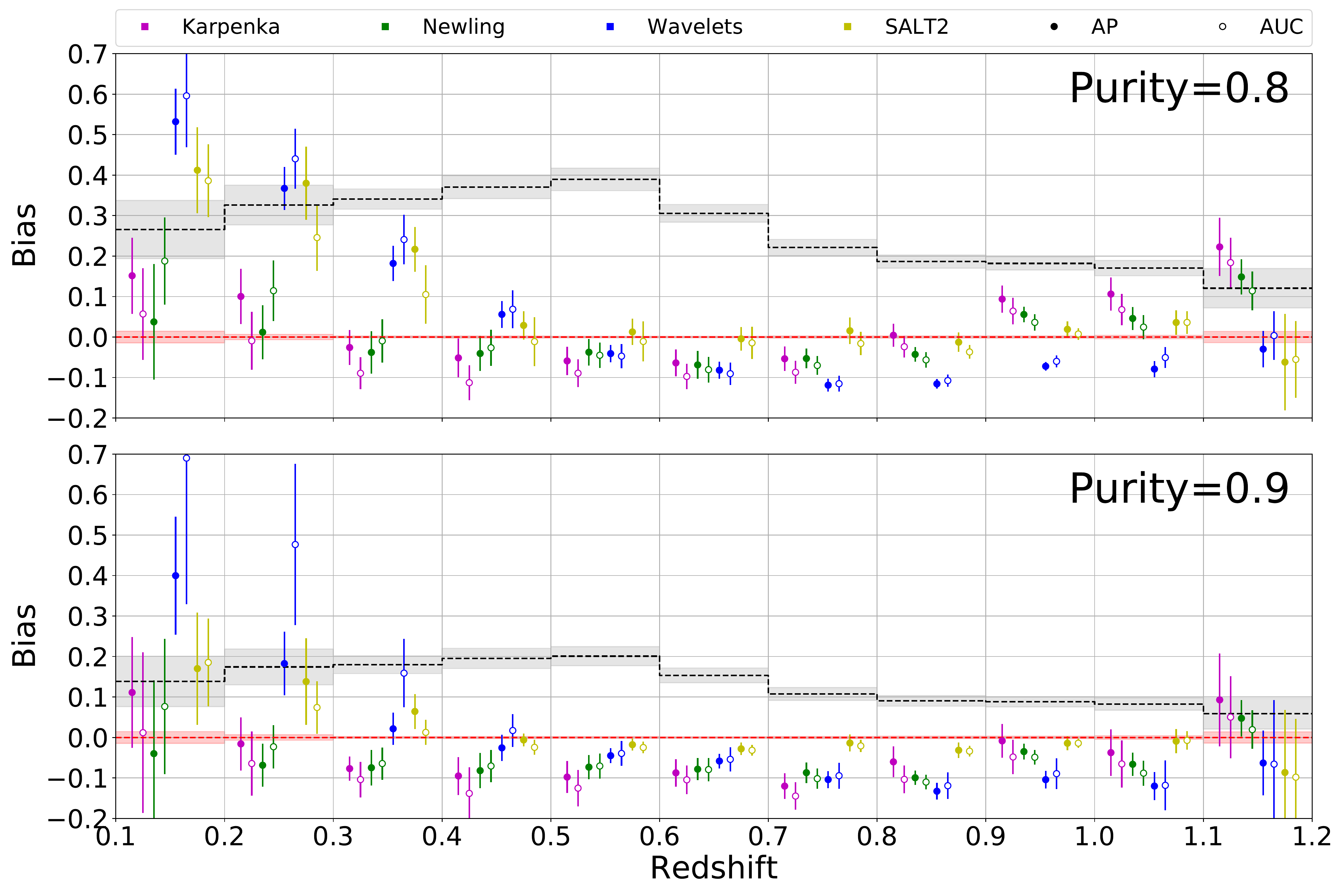}
    \caption{Evolution of bias with the redshift for two arbitrarily chosen values of purity. Data points are slightly displaced left and right for clarity, but are all for the bin centers. The red dashed line and shaded region correspond to values and standard deviation for a perfect classifier with SNeIa only (TP+FN). The black dashed line and gray region represents a completely random classifier. Note that for high purities the bias becomes negative in many cases [see text]. \label{fig:bias}}
\end{figure*}

When we look at the \textit{Completeness-Purity} curves themselves in Figure \ref{fig:prcs} and not just the AP score, we note some important details. When we compare the curves for Newling and Wavelets, both without $z_{\rm host}$ as a feature and looking only at the \tpot\ results we see that if we demand a purity of $95\%$, for instance, Newling reaches a completeness of around $30\%$, significantly larger than that of Wavelets, which lies around $5\%$. Including $z_{\rm host}$ the difference is much smaller but still significant. The final score of Newling is however very similar to the one of Wavelets. This difference in completeness can affect the final cosmological analysis, as we show below. From now on, we focus the analysis only on the catalogs generated by the \tpot\ pipelines with $z_{\rm host}$ as a feature.

\begin{figure*}
    \includegraphics[width=.95\textwidth]{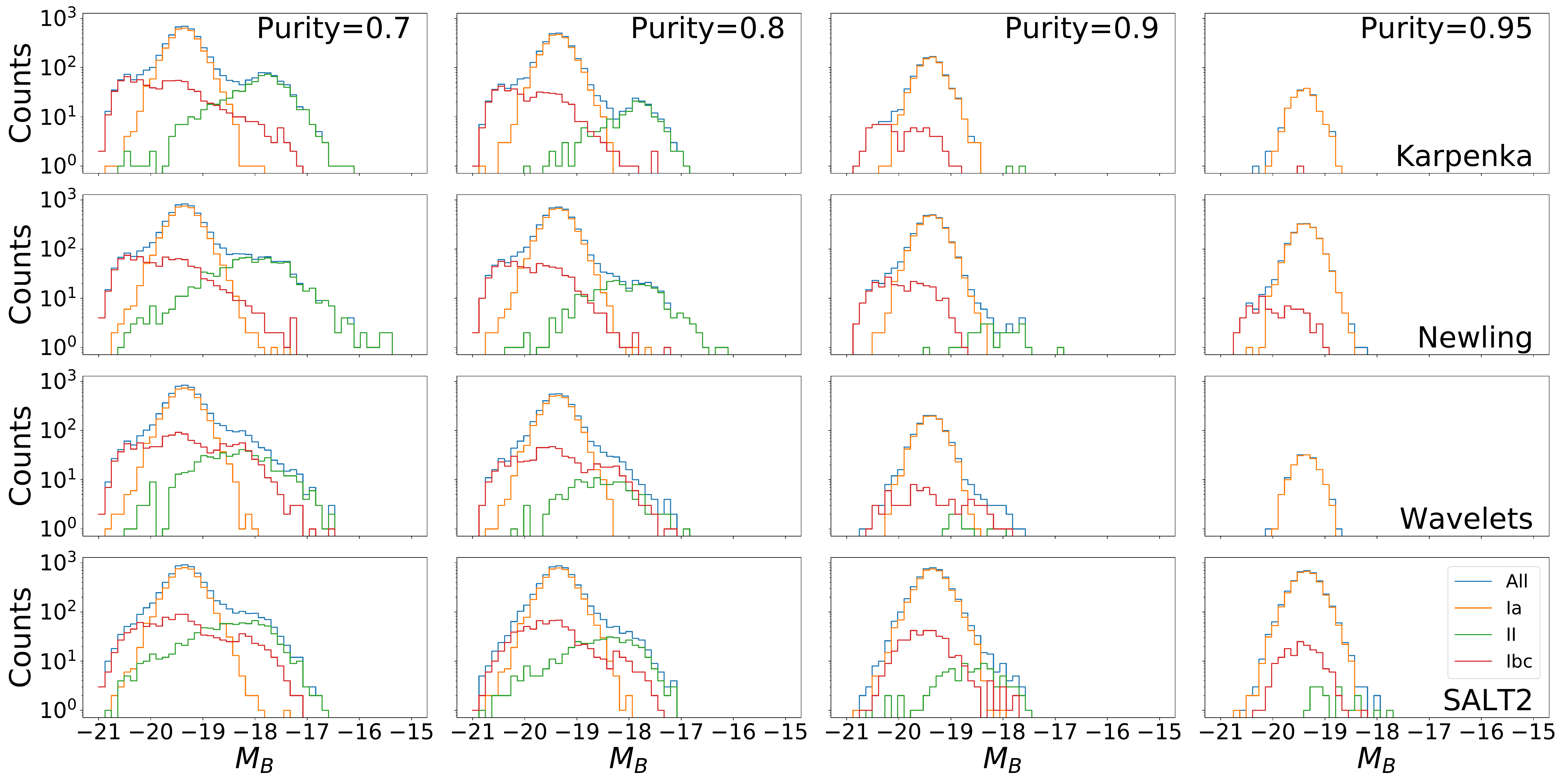}
    \caption{Supernova count by bins of absolute magnitude with AUC as the optimization score. Because the classifier performs best against Type II SNe than against Types Type Ibc, for very high purities only the latter contaminants remain, and because Type II and Types Ibc are at opposite magnitude extremes, the bias
changes sign.}
\label{fig:hists}
\end{figure*}

\begin{table}
\centering
\begin{tabular}{l|c|c}
\toprule
 & AUC & AUC ($+z_{\rm host}$)\\ \hline \hline
This work & 0.969 & 0.985 \\
\cite{Lochner:2016hbn} & 0.969 & 0.979 \\
\cite{Charnock:2016ifh} & 0.901 & 0.940 \\
\cite{Moller2018} & 0.980 & (*)  \\
\bottomrule
\end{tabular}
\caption{Comparison between the best AUC scores in this work and the literature. (*): this information is absent in the original paper. The comparison is made between Accuracy values (defined as $\mathrm{(TP+TN)/N}$)} for the same training set size (1100 SNe), which are equal to $92.4\%$ and $96.8\%$ with and without $z_{host}$, respectively.
\label{tab:scores}
\end{table}

In Table \ref{tab:scores} we show a comparison between the AUC scores, in our best case, and other values quoted in the literature. We focused the comparison only between cases where the authors performed the classification on the SNPCC data set and trained their models with 1100 light curves as in this work. Of the 3 quoted studies, only the one conducted by \Loc~uses the same features as ours. All studies use different classifiers. Only \cite{Moller2018} (using \textit{Recurrent Neural Networks}) reaches higher scores. One advantage of our method is that Decision Tree Ensembles are easier to train than a Neural Network: it takes much less time and can be done by CPUs without the need for GPUs.

\subsection{Analysis of the \emph{Bias-Variance tradeoff}}\label{sec:results_biasvar}

Figure \ref{fig:bias} shows the evolution of bias forall feature sets in comparison to a completely random classifier and a perfect classifier (which returns all SNeIa with $\mathcal{P}=\mathcal{C}=1$), considering two purity cuts (80\% and 90\%). A common feature for all cases is that the bias is low for intermediate redshifts ($0.4 < z < 0.9$), and high for both low and high redshifts ($z < 0.4$, and $z > 0.9$). One may note that the bias values decrease as the purity increases (from the top panel to the bottom one), while the error bars increase due to the loss in the number of classified objects. This general trend does not always hold because the purity in each bin is not necessarily the same, since we fixed only the value of this quantity for the whole sample. The balance is set by the Bias-Variance tradeoff, described in Section \ref{sec:bias_var}. Interestingly though, for high purities the bias becomes negative, so the tradeoff actually penalises very high purities not only due to the low completeness, but also due to the higher absolute values of the bias. One would naively expect the bias to be a monotonic function of the purity, but this a priori unexpected behavior of the bias is explained below.

In Figure \ref{fig:hists} we depict how each SN type contributes to the magnitudes by showing the bin counts of fitted individual absolute magnitudes ($M_{B,i} \equiv m_{B,i}-\mu_{\rm fid}(z_{i, {\rm sim}})$) for four different purity values (70\%, 80\%, 90\% and 95\%). The average of all $M_{B,i}$ is the mean absolute magnitude $M_{B}$ of the SNeIa (see Appendix~\ref{app:salt2}). The overall trend is that the type II SN are the first to be excluded as we increase the purity threshold, followed by the type Ibc. This means that the algorithms are more efficient at distinguishing between types I and II than between Ia and CC. Since types II and Ibc are on opposing $M_B$ extremes, and type II has a larger magnitude range, as the purity threshold increases the magnitude bias due to CC SNe changes sign, instead of monotonically approaching zero. Interestingly, in the case of SALT2 and Wavelets, for large purity thresholds our classifier ends up mis-classifying mostly the CC SNe which have similar magnitudes to the ones of type Ia. This means that for these feature sets the CC contaminants for high purity thresholds may end up mimicking the type Ia standard candles. This peculiar behavior of the bias is fundamentally due to the ML trying to reduce the 8 classes of SN to a simple binary classification: Ia versus CC.

\begin{figure*}
    \includegraphics[width=0.57 \textwidth]{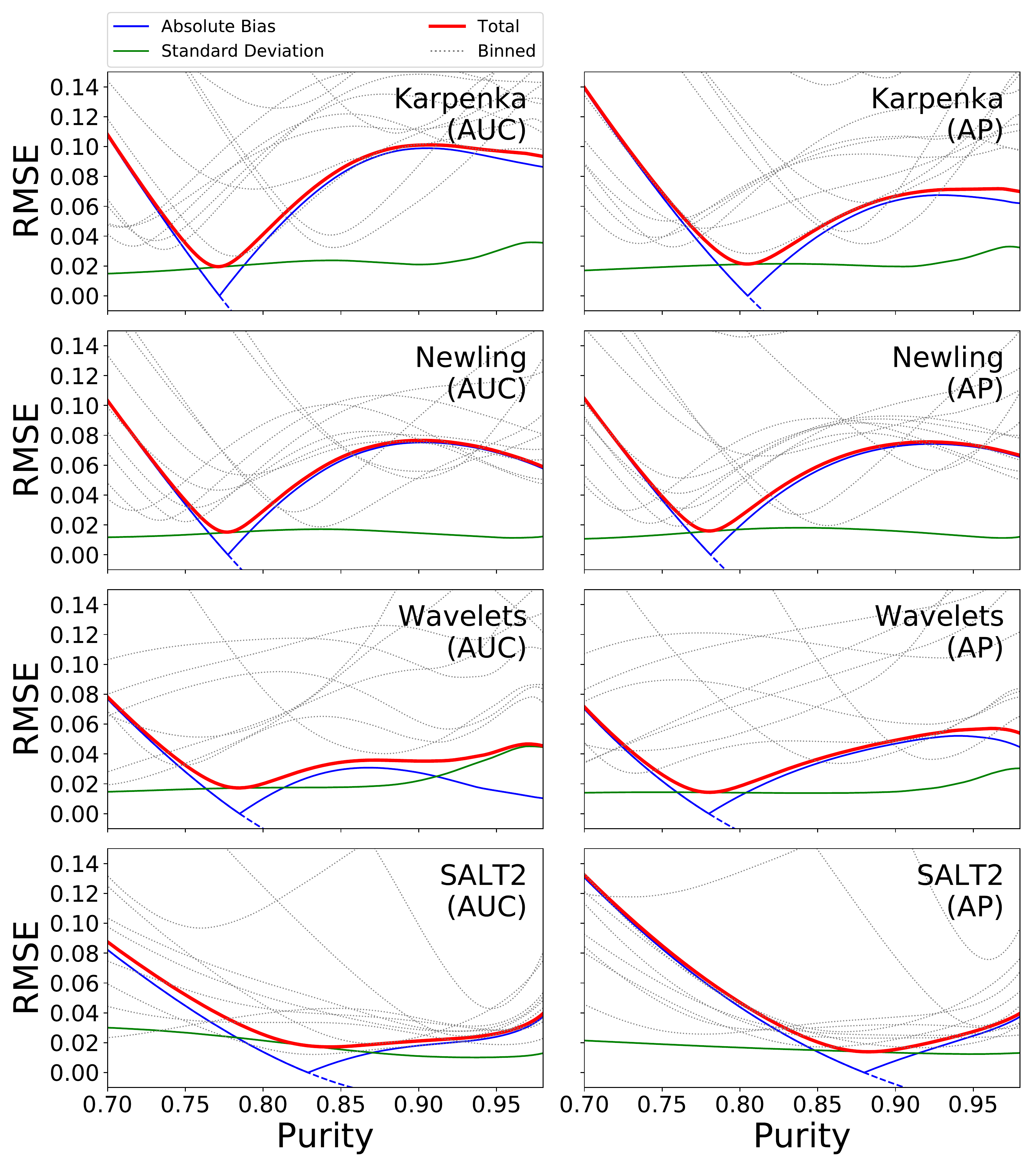}
    \caption{Evolution of RMSE with the Purity, using AUC (left) and AP (right) as scores. The green line is the contribution due to the variance, the blue line is the contribution due to the bias and the red thick line is the total RMSE. The dotted gray lines are the RMSE computed in each redshift bin.
    \label{fig:rmse_tot}}
\end{figure*}

\begin{figure*}
    \includegraphics[width=0.65\textwidth]{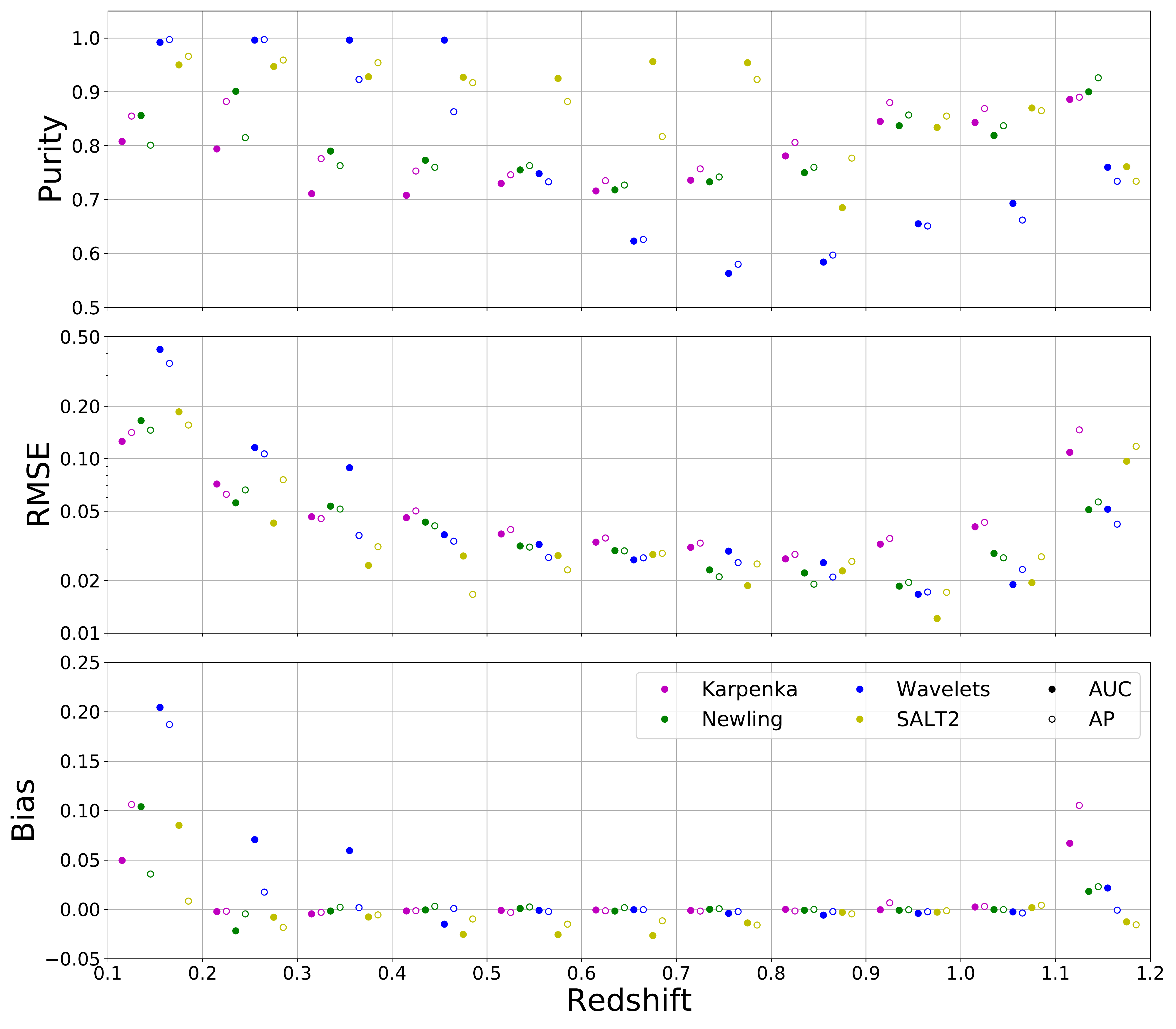}
    \caption{Evolution of purity, RMSE and bias with redshift for the ``binned purity'' case and for all 4 light-curve models considered. \label{fig:rmse-bin}}
\end{figure*}

The dependence of the RMSE with the purity is shown in Figure~\ref{fig:rmse_tot}. The green line is the contribution due to the variance, the blue one is the contribution due to the bias and the red thick line is the total RMSE. The dotted gray lines are the RMSE computed in each redshift bin. Note that, close to the minimum, the variance is almost constant while the bias changes sign. I.e.~the point at which the bias changes sign corresponds approximately to the purity at which RSME reaches a minimum.

We use this dependence to choose what is the best probability threshold, once it is univocally determined by the purity. The simplest choice is the probability which minimizes the total MSE, a constant purity cut, however  we may get very high values of RMSE in some redshift bins. We thus understand that the best way of selecting the sample is to choose the purity values  in each redshift bin that minimize the RMSE, which we dub binned purity. Figure~\ref{fig:rmse-bin} shows explicitly that the bias decreases substantially and becomes smaller than $0.03$ for all cases in a large range of redshifts ($0.4 < z < 1.1$), when we choose this selection criteria. 

We therefore construct a new catalog in which each bin is analyzed enforcing the purity threshold that minimizes the MSE of that bin. To find the best method we compute how much each choice affects the final cosmological constraints. In this context the aim is finding the model that fits the data by minimizing
\begin{eqnarray}
    \chi^2 & = & \big[\bm{\mu}-\mu(\bm{z})\big] \cdot  \bm{C}^{-1} \cdot \big[\bm{\mu}-\mu(\bm{z})\big]^T \nonumber \\ 
    & = & \sum_{i=1}^{N=11}\sum_{j=1}^{N=11} \big[\bar{\mu}_i-\mu(\bar{z}_i)\big] C_{ij}^{-1} \big[\bar{\mu}_j-\mu(\bar{z}_j)\big] \,,
\end{eqnarray}
where $\bm{\mu} = (\bar{\mu}_1, \bar{\mu}_2, ..., \bar{\mu}_{11})$, $\bm{z} = (\bar{z}_1, \bar{z}_2, ..., \bar{z}_{11})$, and $\bm{C}$ is the covariance matrix. Defining the error as $\sigma_i^2 = {\rm MSE}_i = \langle \Delta\bar{\mu}_i^2 \rangle$, where as before $\Delta\bar{\mu} \equiv \bar{\mu}- \bar{\mu}_{fid}(z_{\rm sim})$,
we generalize it for correlated bins as $C_{ij} = \langle \Delta\bar{\mu}_i \Delta\bar{\mu}_j \rangle$ so the covariance for a single catalog is $C_{ij} = \Delta\bar{\mu}_i \Delta\bar{\mu}_j$, therefore
\begin{equation}
    C_{ij} = \left(\dfrac{\Var_i}{N_{SN}^i}\right) \delta_{ij} + \mathrm{b}_i \mathrm{b}_{j} \,.
\end{equation}

\begin{figure*}
	\includegraphics[width=.7\textwidth]{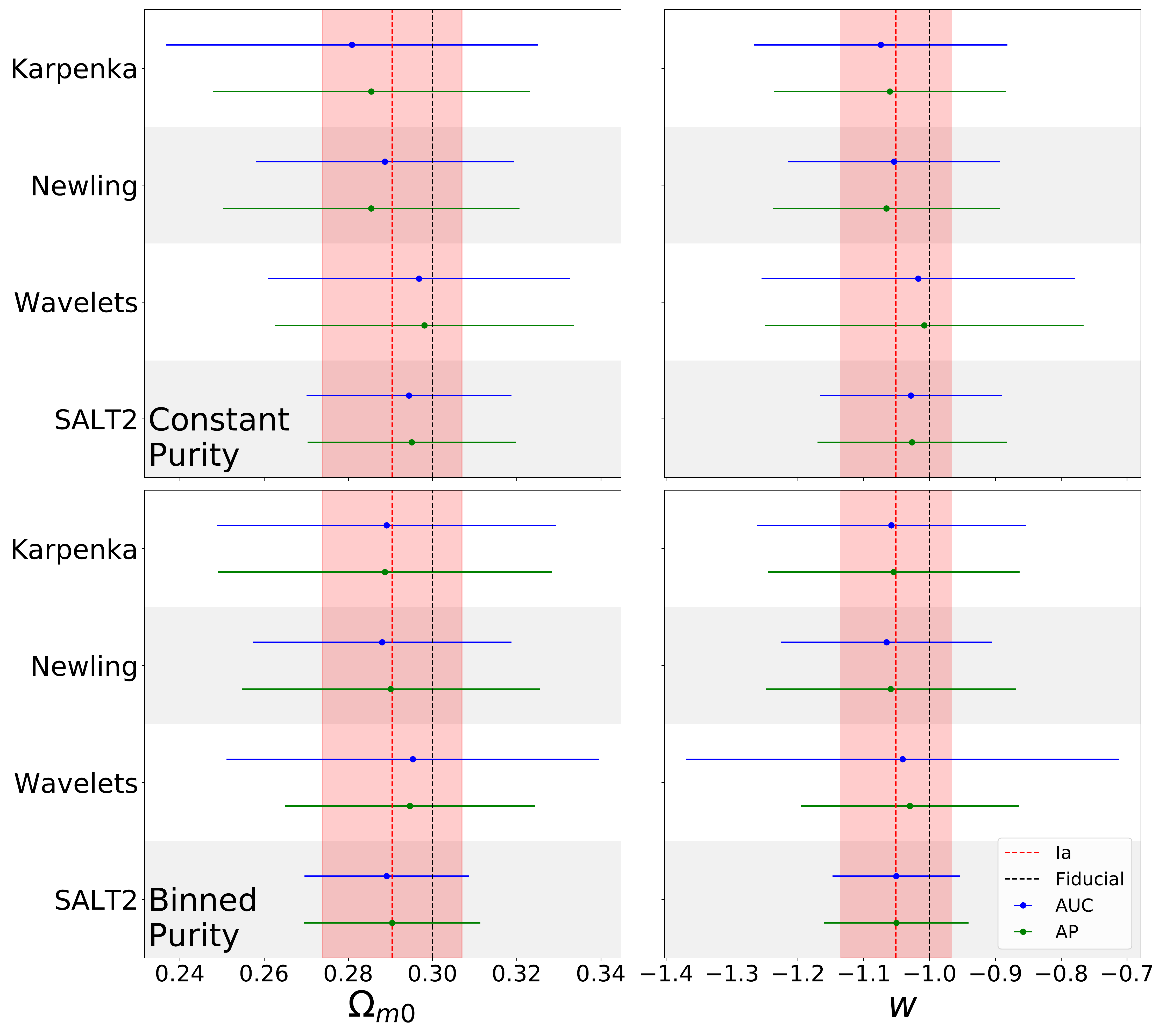}
\caption{$\Omega_{m0}$ and $w$ constraints obtained for each feature set when using a constant purity (top) and choosing purity per redshift bin (bottom). The red shaded region represent the ideal case of perfect classification, for which we only have SNeIa in the final catalog.}
\label{fig:cosmo}
\end{figure*}

{ 
\renewcommand{\arraystretch}{1.5}
\begin{table*}
\centering
\begin{tabular}{lllllllll}
\toprule
{} & \multicolumn{4}{c}{$\Lambda CDM$} & \multicolumn{4}{c}{$wCDM$} \\
{} & \multicolumn{2}{c}{Binned Purity} & \multicolumn{2}{c}{Constant Purity} & \multicolumn{2}{c}{Binned Purity} & \multicolumn{2}{c}{Constant Purity} \\
{} &            AP &         AUC &              AP &         AUC &            AP &         AUC &              AP &         AUC \\
\midrule
Karpenka &     $19\pm 5$ &   $19\pm 6$ &       $20\pm 4$ &   $16\pm 5$ &     $21\pm 6$ &   $19\pm 6$ &       $23\pm 4$ &   $20\pm 5$ \\
Newling  &     $24\pm 7$ &   $32\pm 9$ &       $23\pm 6$ &   $30\pm 4$ &     $22\pm 7$ &  $30\pm 10$ &       $24\pm 5$ &   $27\pm 5$ \\
Wavelets &    $36\pm 13$ &   $16\pm 8$ &       $22\pm 4$ &   $22\pm 4$ &    $30\pm 13$ &   $10\pm 8$ &       $12\pm 3$ &   $13\pm 3$ \\
SALT2    &    $66\pm 15$ &  $75\pm 15$ &      $48\pm 14$ &  $54\pm 17$ &    $62\pm 16$ &  $75\pm 15$ &      $37\pm 12$ &  $44\pm 16$ \\
\bottomrule
\end{tabular}
\caption{The \emph{effective completeness} (in percentage) of each method, for both fitted models. This quantity illustrates how much information is preserved after the photometric classification. The quoted error bars are the standard deviation in this quantity among our 10000 catalogs.}
\label{tab:info}
\end{table*}
}

In order to assess how the different ML techniques behaved in a cosmological analysis, we computed fits for two models: (i) a flat $\Lambda$CDM model with the current matter density parameter $\Omega_{m0}$ as a free parameter; and (ii) a flat $w$CDM model with the dark energy equation of state parameter $w$ running free while fixing $\Omega_{m0}=0.3$. In both cases we marginalize over the overall distance scale which is a combination of the unknown SN absolute magnitude and the Hubble parameter $H_0$.  The constraints on $\Omega_{m0}$ and $w$ are shown in Figure~\ref{fig:cosmo}, where we compare the selection criteria (constant or binned purity thresholds) and the different ML feature sets. We also indicate the result for a perfect classifier (i.e., composed of all the 4800 SNeIa) in red and the fiducial values used in SNPCC ($\Omega_{m0}=0.3$ and $w=-1$) in black. All catalogs classified by ML show small biases when compared to the perfect classifier or the fiducial values (less than half $\sigma$). We also note that  using the binned purity yielded in general slightly improved constraints, but the difference was small. Finally, the SALT2 feature set gives the best results in all cases, approaching the ideal classifier in the binned purity case. On the other hand on average the Karpenka model lead to the worst results, with error bars over twice as large as the ideal case.

We remark that since the small bias on the perfect classifier is smaller than the error bars, it is consistent with random noise in the simulation data. As it is well-known in the literature~\citep{Kessler:2016uwi,Kessler:2018krb}, the combination of lightcurve fitting, selection effects and the intrinsic scatter of SNeIa introduces bias in the cosmological analysis that must be corrected for. Bias removal is thus a very important aspect of supernova cosmology, especially in magnitude-limited surveys. Here however we remove the light-curve fitting biases in a simple way, which we explain in more detail in Appendix~\ref{app:salt2}.

If we assume that all Type Ia supernovae contribute with the same amount of information, the effective statistical error on a cosmological parameter should scale as $\sigma_\mathrm{Ia} = \sigma_\mathrm{I}/\sqrt{N_\mathrm{Ia}}$, where 
$\sigma_\mathrm{I}$ is the standard deviation of the Ia supernovae. We extrapolate on this idea to define $N_\mathrm{Ia}^{\rm eff}$, the effective number of SNeIa in a final photometric classified catalog. The  effective statistical error for this catalog is thus $\sigma = \sigma_\mathrm{I}/\sqrt{N_\mathrm{Ia}^{\rm eff}}$. We likewise define the \emph{effective completeness} $\mathcal{C}^{\rm eff}$ as
\begin{equation}
    \mathcal{C}^{\rm eff} \equiv \frac{N_\mathrm{Ia}^{\rm eff} }{N_\mathrm{Ia}} = \left( \frac{\sigma_{Ia}}{\sigma} \right)^2\,.
\end{equation}
This quantity is computed in Table \ref{tab:info} for different methods. SALT2 preserves up to $75\%$ of the information in the best scenarios, Newling and Wavelets around $30-35\%$ and  Karpenka only $20\%$. This means that even if one drops SALT2 since our SNe may be biased toward it, one still retains around a third of all the information in the SNeIa data without spectroscopy.

To check the validity of our fit we also computed the value of the so-called $R^2$ score (or coefficient of determination\footnote{\url{scikit-learn.org/stable/modules/generated/sklearn.metrics.r2_score.html}}). In all cases $R^2$ was found to be very close to unity ($|1-R^2| \sim 10^{-5}$).

\section{Conclusions}\label{sec:conclusion}

In this work we used simulated data from the Supernova Photometric Classification Challenge (SNPCC) to assess the performance of a series of machine learning methods applied to the problem of supernova classification. We considered some flavors of ensemble Decision Trees: Adaptive Boosting (\textbf{ADA}), RandomForest (\textbf{RF}), ExtraTrees (\textbf{EXT}), Gradient Boosting (\textbf{GB}), and the Extreme Gradient Boosting (\textbf{XGB}) provided by the \textsc{XGBoost} library. We also tested an alternative metric to characterize the performance of the classifier, the Average Purity (AP), and compared the results with the AUC used in \Loc. In addition we performed an automated machine learning training by using \tpot, in order to find the best pipelines for each feature set. Regarding the features, we tested four different models: Karpenka, Newling, wavelets and SALT2. We then explored how a ML-classified SN catalog performs in cosmological parameter estimation, quantifying the degradation due to the imperfect completeness and purity of the resulting catalogs.

Among the feature sets, the SALT2 features gave the best overall results. Considering that this model was used to generate part of the simulated data, it is not clear to what extent the data itself was possibly biased toward this model, and this question certainly deserves to be investigated in more detail in the future. If one thus neglects the SALT2 feature set, both Wavelets and Newling models performed similarly. The Karpenka model performed consistently the worst.

Regarding the scores, we found that the AP was more effective to discriminate between the ML methods as opposed to the AUC, which gave essentially the same results for all boosting methods and \tpot. The boosting methods were considerably better than the averaging ones for both metrics. We showed that, considering the AP score, \tpot~was capable to provide a pipeline more efficient than the boosting methods tested in this work.

We tested two ways for selecting the purity probability threshold: either imposing a constant purity for the whole sample or choosing different purity levels in each redshift bin in order to optimize the MSE. Both choices yielded good results for a flat $\Lambda$CDM model. Employing the Bias-Variance tradeoff we were able to consistently produce photometric classified catalogs that were within half-$\sigma$ of the fiducial model and of a perfect classifier.

One of the most important open questions for future photometric SN surveys is how much cosmological information is preserved after classification. We thus tested this using each feature set with the best ML method, \tpot. We found out that SALT2 is able to preserve up to $75\%$ of the information, while the other second-best alternatives, based on either the Newling model or Wavelet decomposition, around $30-35\%$. This is a very important quantity (it can be interpreted as an effective completeness), as future surveys like LSST will have a huge number of observed SN, but the majority without any spectra. Table~\ref{tab:info} summarizes how much information each fit preserves, compared to a perfect classifier.

It is important to stress that an analysis based on the Bias-Variance Trade-off assumes that the bias cannot be modeled. In fact, as CC SNe lightcurves become better understood it may become possible to predict their impact on the distance estimation and thus mitigate the bias they introduce. This would allow even better performance of the ML photometric classification techniques.

Automated machine learning training techniques are an interesting and very efficient way of attacking the problem of supernova photometric classification for cosmological uses, and we recommend their application in future ML studies in astronomy.

\section*{Acknowledgements}

This work made use of the CHE cluster, managed and funded by COSMO/CBPF/MCTI, with financial support from FINEP and FAPERJ, and of the Milliways cluster at UFRJ, funded by FAPERJ and CNPq. We would like to thank Michelle Lochner for sharing her code, Martin Makler for access to the CHE cluster and Luca Amendola and Luciano Casarini for useful suggestions. MVS is supported by the Brazilian research agency FAPERJ. MQ is also supported by FAPERJ and by the Brazilian research agency CNPq.

\section*{Data availability}
No new data were generated or analysed in support of this research.

\bibliography{bibfile}

\appendix

\section{Distance estimation by SALT2}
\label{app:salt2}

In order to make the residual plots and perform the cosmological constraints we use the SALT2~\citep{Guy:2007dv} as distance estimator, in which the distance modulus for a single supernovae is given by
\begin{equation}\label{eq:mu-salt2}
    \mu = m_B - M_B = m_B^* + \alpha x_1 - \beta c - M_B\,,
\end{equation}
where $m_B^*$, $x_1$ and $c$ are light curve fitting parameters. In particular, $m_B^*$ and $x_0$ are related by
\begin{equation}
m_B^* = -2.5\log_{10}\left[ x_0 \int M_0(t=t_0, \lambda')T^B(\lambda')  d\lambda' \right].
\end{equation}
The parameters $\alpha$, $\beta$ must be fitted along with the cosmological ones. $M_B$ is often left as a free parameter to be marginalized together with $H_0$. In our analysis we fixed the parameters to the fiducial values used in the SNPCC simulations, to wit: $\alpha=0.11$, $\beta=3.2$. We also assumed throughout the simple fiducial model ($\mu_{fid}$) used in SNPCC: flat $\Lambda$CDM with $\Omega_{m0}=0.3$ and $H_0 = 70 \, \mathrm{km/(s \, Mpc)}$.

Based on the assumption that SNeIa are standard candles, $M_B$ in principle should also be a constant and assigned the value used by SNPCC ($M_B=-19.365$). However, as discussed in detail by~\cite{Kessler:2016uwi,Kessler:2018krb}, in any magnitude-limited survey the light-curve fitting itself introduces distance biases, which must be corrected for. These are much larger and independent of usual biases such as the Malmquist bias, and are also independent of selection criteria. The SNPCC data also has additional sources of scatter introduced in the simulations. Together, these effects introduce important biases in the Hubble diagram. 

\begin{figure}
    \includegraphics[width=.49\textwidth]{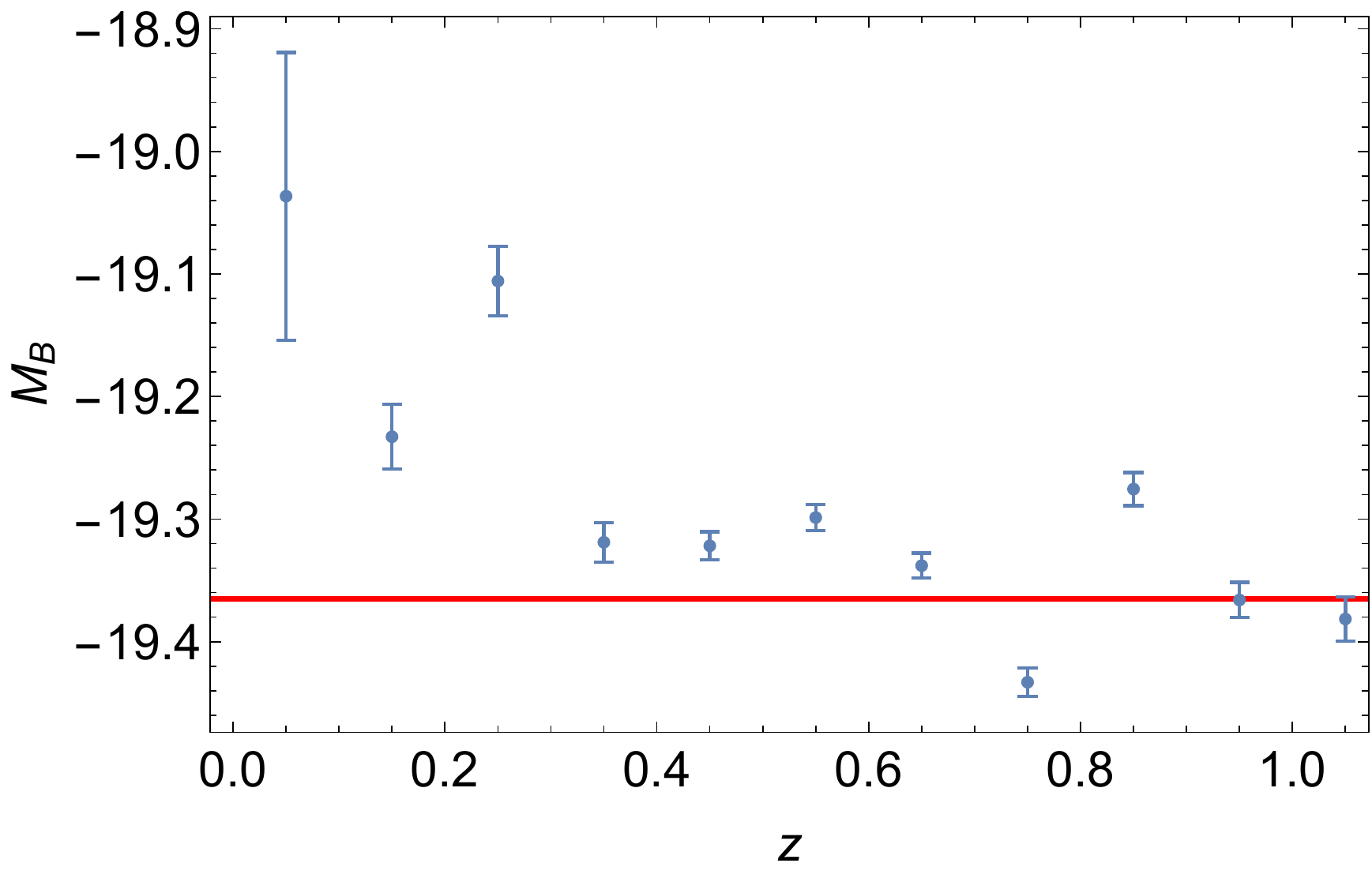}
    \caption{Bestfit values and their standard deviations for the free parameter $M_B$ for the pure type Ia SNPCC catalog. The horizontal line represents the fiducial value used in the simulation. One can see that there are non-negligible sources of bias. Since we are not interested here in the particular details of the SNPCC simulations or in elaborated light-curve fitting bias methods, we remove by hand these biases by adjusting $M_B$ in each redshift bin before implementing the ML methods.}  \label{fig:MB_scatter}
\end{figure}

In this work, as we are not particularly concerned with the usual light-curve fitting biases or the particularities of the SNPCC simulation, we chose not to implement any sophisticated bias removal methods such as the BBC (BEAMS\footnote{BEAMS :
``Bayesian Estimation Applied to Multiple Species'' \citep{Kunz:2006ik}} with Bias Corrections) method of~\cite{Kessler:2016uwi} or a detailed modelling of extra scatter sources. Instead we simply remove by hand \emph{a priori} the bias in $\mu$ by using different values of $M_B$ in each of the $\Delta z=0.1$ redshift bins. This is achieved by making use of the pure SNeIa catalog and the $\mu(z)$ values of the fiducial cosmological model to measure $M_B(z)$ in each redshift bin using~\eqref{eq:mu-salt2}. The difference between the measured $M_B(z)$ value and the fiducial value of $M_B$ used in SNPCC is thus the bias in each bin, which we subsequently remove.
Figure~\ref{fig:MB_scatter} illustrates the fitted values of $M_B$, together with their uncertainties. This procedure guarantees that any cosmological bias in the classified catalogs would be a result of the ML classification code, or of problems in the Bias-Variance tradeoff implementation.

\section{\tpot\ pipelines}
\label{app:tpot}

Here we provide the best pipelines obtained with \tpot\ for each feature set and optimization criterion. They are ready to use as argument of the function \texttt{make\_pipeline} from \texttt{sklearn.pipeline} module. The used functions are from the libraries listed in table~\ref{tab:functions}.

\begin{itemize}
\item Karpenka, optimized by AP:

\begin{lstlisting}
make_union(
    make_pipeline(
        PolynomialFeatures(
            degree=2, include_bias=False,
            interaction_only=False
        ),
        SelectFwe(
            score_func=f_classif,
            alpha=0.033
        )
    ),
    FunctionTransformer(copy)
),
GradientBoostingClassifier(
    learning_rate=0.1, max_depth=9,
    max_features=0.25, min_samples_leaf=8,
    min_samples_split=12, n_estimators=100,
    subsample=0.9
)
\end{lstlisting}

\item Karpenka, optimized by AUC:

\begin{lstlisting}
GradientBoostingClassifier(
    learning_rate=0.1, max_depth=8,
    max_features=0.85, min_samples_leaf=4,
    min_samples_split=11, n_estimators=100,
    subsample=0.9
)
\end{lstlisting}

\item Karpenka $+z_{\rm host}$, optimized by AP:

\begin{lstlisting}
ZeroCount(),
GradientBoostingClassifier(
    learning_rate=0.1, max_depth=10,
    max_features=1.0, min_samples_leaf=8,
    n_estimators=100, subsample=0.75
)
\end{lstlisting}

\item Karpenka $+z_{\rm host}$, optimezed by AUC:

\begin{lstlisting}
make_union(
    FunctionTransformer(copy),
    make_pipeline(
        VarianceThreshold(threshold=0.2),
        FastICA(tol=0.0)
    )
),
XGBClassifier(
    learning_rate=0.1, max_depth=9,
    min_child_weight=2, n_estimators=100,
    nthread=1, subsample=0.8
)
\end{lstlisting}

\item Newling, optimized by AP:

\begin{lstlisting}
make_union(
    make_union(
        FunctionTransformer(copy),
        FunctionTransformer(copy)
    ),
    make_pipeline(
        Normalizer(norm="l2"),
        RFE(estimator=ExtraTreesClassifier(
            criterion="entropy",
            max_features=0.75,
            n_estimators=100
            ), step=0.15
        ),
        RFE(
            estimator=ExtraTreesClassifier(
                criterion="gini",
                max_features=0.05,
                n_estimators=100
            ),
            step=0.95
        ),
        FastICA(tol=0.55),
        RBFSampler(gamma=0.8),
        RFE(
            estimator=ExtraTreesClassifier(
                criterion="entropy",
                max_features=0.15,
                n_estimators=100
            ),
            step=0.4
        )
    )
),
ExtraTreesClassifier(
    bootstrap=False, criterion="entropy",
    max_features=0.95, min_samples_leaf=2,
    min_samples_split=5, n_estimators=100
)
\end{lstlisting}

\item Newling, optimized by AUC:

\begin{lstlisting}
Normalizer(norm="l2"),
StackingEstimator(
    estimator=LinearSVC(
        C=5.0, dual=False,
        loss="squared_hinge",
        penalty="l1", tol=0.01
    )
),
GradientBoostingClassifier(
    learning_rate=0.1, max_depth=9,
    max_features=0.85, min_samples_leaf=14,
    min_samples_split=9, n_estimators=100,
    subsample=0.85
)
\end{lstlisting}

\item Newling $+z_{\rm host}$, optimized by AP:

\begin{lstlisting}
make_union(
    FunctionTransformer(copy),
    FunctionTransformer(copy)
),
MinMaxScaler(),
RFE(
    estimator=ExtraTreesClassifier(
        criterion="entropy",
        max_features=0.3,
        n_estimators=100
    ),
    step=1.0
),
PolynomialFeatures(
    degree=2, include_bias=False,
    interaction_only=False
),
SelectFwe(
    score_func=f_classif, alpha=0.004
),
ExtraTreesClassifier(
    bootstrap=False, criterion="entropy",
    max_features=0.25, min_samples_leaf=2,
    min_samples_split=3, n_estimators=100
)
\end{lstlisting}

\item Newling $+z_{\rm host}$, optimized by AUC:

\begin{lstlisting}
RFE(
    estimator=ExtraTreesClassifier(
        criterion="entropy",
        max_features=0.75,
        n_estimators=100
    ),
    step=0.2
),
ExtraTreesClassifier(
    bootstrap=False,
    criterion="gini",
    max_features=0.95,
    min_samples_leaf=1,
    min_samples_split=2,
    n_estimators=100
)
\end{lstlisting}

\item SALT2, optimized by AP:

\begin{lstlisting}
StackingEstimator(
    estimator=LinearSVC(
        C=20.0, dual=False,
        loss="squared_hinge",
        penalty="l2", tol=0.0001
    )
),
StackingEstimator(
    estimator=LinearSVC(
        C=10.0, dual=True,
        loss="squared_hinge",
        penalty="l2", tol=0.1
    )
),
ZeroCount(),
RandomForestClassifier(
    bootstrap=False, criterion="entropy",
    max_features=0.35, min_samples_leaf=3,
    min_samples_split=9, n_estimators=100
)
\end{lstlisting}

\item SALT2, optimized by AUC:

\begin{lstlisting}
make_union(
    FunctionTransformer(copy),
    make_union(
        make_union(
            make_pipeline(
                RBFSampler(gamma=0.35),
                Normalizer(norm="l1")
            ),
            FunctionTransformer(copy)
        ),
        FunctionTransformer(copy)
    )
),
RandomForestClassifier(
    bootstrap=False, criterion="entropy",
    max_features=0.2, min_samples_leaf=2,
    min_samples_split=7, n_estimators=100
)
\end{lstlisting}

\item SALT2 $+z_{\rm host}$, optimized by AP:

\begin{lstlisting}
make_union(
    FunctionTransformer(copy),
    make_union(
        make_pipeline(
            FastICA(tol=0.65),
            FastICA(tol=0.0),
            ZeroCount()
        ),
        FunctionTransformer(copy)
    )
),
RandomForestClassifier(
    bootstrap=False, criterion="entropy",
    max_features=0.4, min_samples_leaf=1,
    min_samples_split=4, n_estimators=100
)
\end{lstlisting}

\item SALT2 $+z_{\rm host}$, optimized by AUC:

\begin{lstlisting}
make_union(
    FunctionTransformer(copy),
    FastICA(tol=0.05)
),
StackingEstimator(
    estimator=LogisticRegression(
        C=0.5, dual=False,
        penalty="l1"
    )
),
RandomForestClassifier(
    bootstrap=True, criterion="entropy",
    max_features=0.2, min_samples_leaf=1,
    min_samples_split=2, n_estimators=100
)
\end{lstlisting}

\item Wavelets, optimized by AP:

\begin{lstlisting}
PolynomialFeatures(
    degree=2, include_bias=False,
    interaction_only=False
),
StackingEstimator(
    estimator=ExtraTreesClassifier(
        bootstrap=True,
        criterion="gini",
        max_features=0.25,
        min_samples_leaf=7,
        min_samples_split=17,
        n_estimators=100
    )
),
StackingEstimator(
    estimator=LogisticRegression(
        C=1.0, dual=False, penalty="l1"
    )
),
MaxAbsScaler(),
RFE(
    estimator=ExtraTreesClassifier(
        criterion="entropy",
        max_features=0.75,
        n_estimators=100
    ),
    step=0.1
),
StackingEstimator(estimator=GaussianNB()),
MinMaxScaler(),
XGBClassifier(
    learning_rate=0.1, max_depth=4,
    min_child_weight=1, n_estimators=100,
    nthread=1, subsample=0.6
)
\end{lstlisting}

\item Wavelets, optimized by AUC:

\begin{lstlisting}
make_union(
    make_pipeline(
        Normalizer(norm="l2"),
        FastICA(tol=0.7)
    ),
    FunctionTransformer(copy)
),
StackingEstimator(
    estimator=ExtraTreesClassifier(
        bootstrap=False,
        criterion="entropy",
        max_features=0.9,
        min_samples_leaf=3,
        min_samples_split=6,
        n_estimators=100
    )
),
RobustScaler(),
ZeroCount(),
LogisticRegression(
    C=0.5, dual=True,
    penalty="l2"
)
\end{lstlisting}

\item Wavelets $+z_{\rm host}$, optimized by AP:

\begin{lstlisting}
make_union(
    Nystroem(
        gamma=0.15, kernel="linear",
        n_components=10
    ),
    StackingEstimator(
        estimator=make_pipeline(
            StackingEstimator(
                estimator=LogisticRegression(
                    C=25.0,
                    dual=False,
                    penalty="l1"
                )
            ),
            PolynomialFeatures(
                degree=2,
                include_bias=False,
                interaction_only=False
            ),
            LogisticRegression(
                C=0.001, dual=False,
                penalty="l2"
            )
        )
    )
),
GradientBoostingClassifier(
    learning_rate=0.1, max_depth=8,
    max_features=0.8, min_samples_leaf=4,
    min_samples_split=10, n_estimators=100,
    subsample=0.9
)
\end{lstlisting}

\item Wavelets $+z_{\rm host}$, optimized by AUC:

{ 
\renewcommand{\arraystretch}{1.5}
\begin{table*}
\centering
\begin{tabular}{|l|l|l|l|}
 \hline \textbf{Library} & \textbf{Module} & \textbf{Function} & \textbf{Website}\\ \hline \hline
 \texttt{copy} & & \texttt{copy} & \url{docs.python.org/3/library/copy.html}\\ \hline
 \texttt{xgboost} & & \texttt{XGBClassifier} & \url{xgboost.readthedocs.io} \\ \hline
 \multirow{2}{*}{\texttt{tpot}} & \multirow{2}{*}{\texttt{builtins}} & \texttt{StackingEstimator} & \multirow{2}{*}{\url{epistasislab.github.io/tpot}}\\
 & & \texttt{ZeroCount} & \\  \hline
 \multirow{21}{*}{\texttt{sklearn}} & \texttt{pipeline} & \texttt{make\_union} & \multirow{21}{*}{\url{scikit-learn.org}} \\ \cline{2-3}
 & \texttt{decomposition} & \texttt{FastICA} & \\ \cline{2-3}
 & \texttt{linear\_model} & \texttt{LogisticRegression} & \\ \cline{2-3}
 & \texttt{svm} & \texttt{LinearSVC} & \\ \cline{2-3}
 & \texttt{neighbors} & \texttt{KNeighborsClassifier} & \\ \cline{2-3}
 & \texttt{naive\_bayes} & \texttt{GaussianNB} & \\ \cline{2-3}
 & \multirow{2}{*}{\texttt{kernel\_approximation}} & \texttt{Nystroem} & \\ 
 & & \texttt{RBFSampler} & \\ \cline{2-3}
 & \multirow{3}{*}{\texttt{ensemble}} & \texttt{ExtraTreesClassifier} & \\
 & & \texttt{GradientBoostingClassifier} & \\
 & & \texttt{RandomForestClassifier} & \\ \cline{2-3}
 & \multirow{4}{*}{\texttt{feature\_selection}} & \texttt{RFE} & \\
 & & \texttt{SelectFwe} & \\
 & & \texttt{f\_classif} & \\
 & & \texttt{VarianceThreshold} & \\ \cline{2-3}
 & \multirow{6}{*}{\texttt{preprocessing}} & \texttt{FunctionTransformer} & \\
 & & \texttt{MaxAbsScaler} & \\
 & & \texttt{MinMaxScaler} & \\
 & & \texttt{PolynomialFeatures} & \\
 & & \texttt{Normalizer} & \\
 & & \texttt{RobustScaler} & \\ \hline
\end{tabular}
\caption{Functions used in pipelines built by the \texttt{tpot} genetic optimization.}\label{tab:functions}
\end{table*}
}

\begin{lstlisting}
make_union(
    StackingEstimator(
        estimator=KNeighborsClassifier(
            n_neighbors=21, p=1,
            weights="uniform"
        )
    ),
    StackingEstimator(
        estimator=make_pipeline(
            RBFSampler(gamma=0.1),
            ExtraTreesClassifier(
                bootstrap=False,
                criterion="entropy",
                max_features=0.25,
                min_samples_leaf=18,
                min_samples_split=7,
                n_estimators=100
            )
        )
    )
),
MaxAbsScaler(),
GradientBoostingClassifier(
    learning_rate=0.1, max_depth=7,
    max_features=0.9, min_samples_leaf=12,
    min_samples_split=9, n_estimators=100,
    subsample=0.85
)
\end{lstlisting}
\end{itemize}

\bsp	
\label{lastpage}
\end{document}